\documentclass[12pt]{article}
\usepackage{amsmath, amssymb, graphics}
\usepackage{amsbsy}
\usepackage{amsfonts}
\usepackage{amsmath}
\usepackage{graphicx}
\usepackage{array}
\newcommand{\mathsym}[1]{{}}
\newcommand{\ket}[1]{\left\lvert #1 \right\rangle}
\renewcommand{\title}[1]{\vbox{\center\LARGE{#1}}\vspace{5mm}}
\renewcommand{\author}[1]{\vbox{\center#1}\vspace{5mm}}

\makeatletter
\renewcommand\section{\@startsection {section}{1}{\z@}%
                                   {-3.5ex \@plus -1ex \@minus -.2ex}%
                                   {2.3ex \@plus.2ex}%
                                   {\normalfont\large\bfseries}}
\renewcommand\subsection{\@startsection{subsection}{2}{\z@}%
                                   {-3.25ex\@plus -1ex \@minus -.2ex}%
                                   {1.5ex \@plus .2ex}%
                                   {\normalfont\normalsize\bfseries}}
\makeatother

\renewcommand{\theequation}{1.\arabic{equation}}

\renewcommand{\[}{\begin{eqnarray}}
\renewcommand{\]}{\end{eqnarray}}
\newcommand{\nn}{\nonumber}

\newcommand{\OS}{$OSp(8^{*}|4)$}

\newcommand{\p}{{\partial}}

\def\moth{\mathsurround=0pt}
\newdimen\zo \zo=0pt

\def\tick{\leaders\hrule height 0.5ex depth 0pt \hskip 0.5pt}
\def\upboxfill{$\moth \setbox\zo\hbox{\tick}%
  \hskip 2pt\hbox to 0pt{$\tick$\hss}\hrulefill \hbox to 2pt{$\tick$\hss}$}

\def\dtick{\leaders\hrule height .34pt depth 0.5ex \hskip 0.5pt}
\def\downboxfill{$\moth \setbox\zo\hbox{\dtick}%
  \hskip 2pt\hbox to 0pt{$\dtick$\hss}\hrulefill%
  \hbox to 2pt{$\dtick$\hss}$}

\newcommand{\stln}{\setlength{\unitlength}{2.2ex}}
\newcommand{\fr}{\framebox(1,1){}}
\newcommand{\sfr}{\framebox(1,1){\begin{picture}(1,1)
  \put(0,0){\line(1,1){1}}\end{picture}}}

\newcommand{\onebox}
{\stln \lower1.4ex\hbox{
\begin{picture}(1.6,1.6)
\put(.3,.3){\fr}
\end{picture}}}

\newcommand{\twobox}
{\stln \lower1.4ex\hbox{
\begin{picture}(2.6,1.6)
\put(.3,.3){\fr}
\put(1.3,.3){\fr}
\end{picture}}}

\newcommand{\threebox}
{\stln \lower1.4ex\hbox{
\begin{picture}(3.6,1.6)
\multiput(.3,.3)(1,0){3}{\fr}
\end{picture}}}

\newcommand{\fourbox}
{\stln \lower1.4ex\hbox{
\begin{picture}(4.6,1.6)
\multiput(.3,.3)(1,0){4}{\fr}
\end{picture}}}

\newcommand{\fivebox}
{\stln \lower1.4ex\hbox{
\begin{picture}(5.6,1.6)
\multiput(.3,.3)(1,0){5}{\fr}
\end{picture}}}

\newcommand{\sixbox}
{\stln \lower1.4ex\hbox{
\begin{picture}(6.6,1.6)
\multiput(.3,.3)(1,0){6}{\fr}
\end{picture}}}

\newcommand{\genrowbox}
{\stln \lower1.4ex\hbox{
\begin{picture}(7.6,1.6)
\multiput(.3,.3)(1,0){3}{\fr}
\put(3.3,.3){\framebox(3,1){$\cdots$}}
\put(6.3,.3){\fr}
\end{picture}}}

\newcommand{\oneonebox}
{\stln \lower2.6ex\hbox{
\begin{picture}(1.6,2.6)
\put(.3,.3){\fr}
\put(.3,1.3){\fr}
\end{picture}}}

\newcommand{\twoonebox}
{\stln \lower2.6ex\hbox{
\begin{picture}(2.6,2.6)
\put(.3,1.3){\fr}
\put(1.3,1.3){\fr}
\put(0.3,0.3){\fr}
\end{picture}}}

\newcommand{\threeonebox}
{\stln \lower2.6ex\hbox{
\begin{picture}(3.6,2.6)
\multiput(.3,1.3)(1,0){3}{\fr}
\put(.3,.3){\fr}
\end{picture}}}

\newcommand{\fouronebox}
{\stln \lower2.6ex\hbox{
\begin{picture}(4.6,2.6)
\multiput(.3,1.3)(1,0){4}{\fr}
\put(.3,.3){\fr}
\end{picture}}}

\newcommand{\fiveonebox}
{\stln \lower2.6ex\hbox{
\begin{picture}(5.6,2.6)
\multiput(.3,1.3)(1,0){5}{\fr}
\put(.3,.3){\fr}
\end{picture}}}

\newcommand{\sixonebox}
{\stln \lower2.6ex\hbox{
\begin{picture}(6.6,2.6)
\multiput(.3,1.3)(1,0){6}{\fr}
\put(.3,.3){\fr}
\end{picture}}}

\newcommand{\twotwobox}
{\stln \lower2.6ex\hbox{
\begin{picture}(2.6,2.6)
\put(.3,.3){\fr}
\put(.3,1.3){\fr}
\put(1.3,.3){\fr}
\put(1.3,1.3){\fr}
\end{picture}}}

\newcommand{\threetwobox}
{\stln \lower2.6ex\hbox{
\begin{picture}(3.6,2.6)
\multiput(.3,1.3)(1,0){3}{\fr}
\put(.3,.3){\fr}
\put(1.3,.3){\fr}
\end{picture}}}

\newcommand{\fourtwobox}
{\stln \lower2.6ex\hbox{
\begin{picture}(4.6,2.6)
\multiput(.3,1.3)(1,0){4}{\fr}
\put(.3,.3){\fr}
\put(1.3,.3){\fr}
\end{picture}}}

\newcommand{\fivetwobox}
{\stln \lower2.6ex\hbox{
\begin{picture}(5.6,2.6)
\multiput(.3,1.3)(1,0){5}{\fr}
\put(.3,.3){\fr}
\put(1.3,.3){\fr}
\end{picture}}}

\newcommand{\sixtwobox}
{\stln \lower2.6ex\hbox{
\begin{picture}(6.6,2.6)
\multiput(.3,1.3)(1,0){6}{\fr}
\put(.3,.3){\fr}
\put(1.3,.3){\fr}
\end{picture}}}

\newcommand{\threethreebox}
{\stln \lower2.6ex\hbox{
\begin{picture}(3.6,2.6)
\multiput(.3,1.3)(1,0){3}{\fr}
\multiput(.3,.3)(1,0){3}{\fr}
\end{picture}}}

\newcommand{\fourthreebox}
{\stln \lower2.6ex\hbox{
\begin{picture}(4.6,2.6)
\multiput(.3,1.3)(1,0){4}{\fr}
\multiput(.3,.3)(1,0){3}{\fr}
\end{picture}}}

\newcommand{\fivethreebox}
{\stln \lower2.6ex\hbox{
\begin{picture}(5.6,2.6)
\multiput(.3,1.3)(1,0){5}{\fr}
\multiput(.3,.3)(1,0){3}{\fr}
\end{picture}}}

\newcommand{\sixthreebox}
{\stln \lower2.6ex\hbox{
\begin{picture}(6.6,2.6)
\multiput(.3,1.3)(1,0){6}{\fr}
\multiput(.3,.3)(1,0){3}{\fr}
\end{picture}}}

\newcommand{\fourfourbox}
{\stln \lower2.6ex\hbox{
\begin{picture}(4.6,2.6)
\multiput(.3,1.3)(1,0){4}{\fr}
\multiput(.3,.3)(1,0){4}{\fr}
\end{picture}}}

\newcommand{\fivefourbox}
{\stln \lower2.6ex\hbox{
\begin{picture}(5.6,2.6)
\multiput(.3,1.3)(1,0){5}{\fr}
\multiput(.3,.3)(1,0){4}{\fr}
\end{picture}}}

\newcommand{\sixfourbox}
{\stln \lower2.6ex\hbox{
\begin{picture}(6.6,2.6)
\multiput(.3,1.3)(1,0){6}{\fr}
\multiput(.3,.3)(1,0){4}{\fr}
\end{picture}}}

\newcommand{\fivefivebox}
{\stln \lower2.6ex\hbox{
\begin{picture}(5.6,2.6)
\multiput(.3,1.3)(1,0){5}{\fr}
\multiput(.3,.3)(1,0){5}{\fr}
\end{picture}}}

\newcommand{\sixfivebox}
{\stln \lower2.6ex\hbox{
\begin{picture}(6.6,2.6)
\multiput(.3,1.3)(1,0){6}{\fr}
\multiput(.3,.3)(1,0){5}{\fr}
\end{picture}}}

\newcommand{\sixsixbox}
{\stln \lower2.6ex\hbox{
\begin{picture}(6.6,2.6)
\multiput(.3,1.3)(1,0){6}{\fr}
\multiput(.3,.3)(1,0){6}{\fr}
\end{picture}}}

\newcommand{\oneoneonebox}
{\stln \lower3.8ex\hbox{
\begin{picture}(1.6,3.6)
\multiput(.3,.3)(0,1){3}{\fr}
\end{picture}}}

\newcommand{\twooneonebox}
{\stln \lower3.8ex\hbox{
\begin{picture}(2.6,3.6)
\multiput(.3,.3)(0,1){3}{\fr}
\put(1.3,2.3){\fr}
\end{picture}}}

\newcommand{\twotwoonebox}
{\stln \lower3.8ex\hbox{
\begin{picture}(2.6,3.6)
\multiput(.3,.3)(0,1){3}{\fr}
\put(1.3,1.3){\fr}
\put(1.3,2.3){\fr}
\end{picture}}}

\newcommand{\twotwotwobox}
{\stln \lower3.8ex\hbox{
\begin{picture}(2.6,3.6)
\multiput(.3,.3)(0,1){3}{\fr}
\multiput(1.3,.3)(0,1){3}{\fr}
\end{picture}}}

\newcommand{\threeoneonebox}
{\stln \lower3.8ex\hbox{
\begin{picture}(3.6,3.6)
\multiput(.3,.3)(0,1){3}{\fr}
\put(1.3,2.3){\fr}
\put(2.3,2.3){\fr}
\end{picture}}}

\newcommand{\threeoneoneonebox}
{\stln \lower3.8ex\hbox{
\begin{picture}(3.6,3.6)
\multiput(.3,.3)(0,1){3}{\fr}
\put(1.3,2.3){\fr}
\put(2.3,2.3){\fr}
\put(3.3,2.3){\fr}
\end{picture}}}

\newcommand{\threetwoonebox}
{\stln \lower3.8ex\hbox{
\begin{picture}(3.6,3.6)
\multiput(.3,.3)(0,1){3}{\fr}
\put(1.3,2.3){\fr}
\put(2.3,2.3){\fr}
\put(1.3,1.3){\fr}
\end{picture}}}

\newcommand{\threetwotwobox}
{\stln \lower3.8ex\hbox{
\begin{picture}(3.6,3.6)
\multiput(.3,.3)(0,1){3}{\fr}
\multiput(1.3,.3)(0,1){3}{\fr}
\put(2.3,2.3){\fr}
\end{picture}}}

\newcommand{\threethreeonebox}
{\stln \lower3.8ex\hbox{
\begin{picture}(3.6,3.6)
\multiput(.3,2.3)(1,0){3}{\fr}
\multiput(.3,1.3)(1,0){3}{\fr}
\put(.3,.3){\fr}
\end{picture}}}

\newcommand{\threethreetwobox}
{\stln \lower3.8ex\hbox{
\begin{picture}(3.6,3.6)
\multiput(.3,2.3)(1,0){3}{\fr}
\multiput(.3,1.3)(1,0){3}{\fr}
\put(.3,.3){\fr}
\put(1.3,.3){\fr}
\end{picture}}}

\newcommand{\threethreethreebox}
{\stln \lower3.8ex\hbox{
\begin{picture}(3.6,3.6)
\multiput(.3,2.3)(1,0){3}{\fr}
\multiput(.3,1.3)(1,0){3}{\fr}
\multiput(.3,.3)(1,0){3}{\fr}
\end{picture}}}

\newcommand{\gencolbox}
{\stln \lower8.6ex\hbox{
\begin{picture}(1.6,7.6)
\multiput(.3,4.3)(0,1){3}{\fr}
\put(.3,1.3){\framebox(1,3){$\vdots$}}
\put(.3,.3){\fr}
\end{picture}}}

\newcommand{\sonebox}
{\stln \lower1.4ex\hbox{
\begin{picture}(1.6,1.6)
\put(.3,.3){\sfr}
\end{picture}}}

\newcommand{\stwobox}
{\stln \lower1.4ex\hbox{
\begin{picture}(2.6,1.6)
\put(.3,.3){\sfr}
\put(1.3,.3){\sfr}
\end{picture}}}

\newcommand{\sthreebox}
{\stln \lower1.4ex\hbox{
\begin{picture}(3.6,1.6)
\multiput(.3,.3)(1,0){3}{\sfr}
\end{picture}}}

\newcommand{\sfourbox}
{\stln \lower1.4ex\hbox{
\begin{picture}(4.6,1.6)
\multiput(.3,.3)(1,0){4}{\sfr}
\end{picture}}}

\newcommand{\sfivebox}
{\stln \lower1.4ex\hbox{
\begin{picture}(5.6,1.6)
\multiput(.3,.3)(1,0){5}{\sfr}
\end{picture}}}

\newcommand{\ssixbox}
{\stln \lower1.4ex\hbox{
\begin{picture}(6.6,1.6)
\multiput(.3,.3)(1,0){6}{\sfr}
\end{picture}}}

\newcommand{\sgenrowbox}
{\stln \lower1.4ex\hbox{
\begin{picture}(7.6,1.6)
\multiput(.3,.3)(1,0){3}{\sfr}
\put(3.3,.3){\framebox(3,1){$\cdots$}}
\put(6.3,.3){\sfr}
\end{picture}}}

\newcommand{\soneonebox}
{\stln \lower2.6ex\hbox{
\begin{picture}(1.6,2.6)
\put(.3,.3){\sfr}
\put(.3,1.3){\sfr}
\end{picture}}}

\newcommand{\stwoonebox}
{\stln \lower2.6ex\hbox{
\begin{picture}(2.6,2.6)
\put(.3,1.3){\sfr}
\put(1.3,1.3){\sfr}
\put(0.3,0.3){\sfr}
\end{picture}}}

\newcommand{\sthreeonebox}
{\stln \lower2.6ex\hbox{
\begin{picture}(3.6,2.6)
\multiput(.3,1.3)(1,0){3}{\sfr}
\put(.3,.3){\sfr}
\end{picture}}}

\newcommand{\sfouronebox}
{\stln \lower2.6ex\hbox{
\begin{picture}(4.6,2.6)
\multiput(.3,1.3)(1,0){4}{\sfr}
\put(.3,.3){\sfr}
\end{picture}}}

\newcommand{\sfiveonebox}
{\stln \lower2.6ex\hbox{
\begin{picture}(5.6,2.6)
\multiput(.3,1.3)(1,0){5}{\sfr}
\put(.3,.3){\sfr}
\end{picture}}}

\newcommand{\ssixonebox}
{\stln \lower2.6ex\hbox{
\begin{picture}(6.6,2.6)
\multiput(.3,1.3)(1,0){6}{\sfr}
\put(.3,.3){\sfr}
\end{picture}}}

\newcommand{\stwotwobox}
{\stln \lower2.6ex\hbox{
\begin{picture}(2.6,2.6)
\put(.3,.3){\sfr}
\put(.3,1.3){\sfr}
\put(1.3,.3){\sfr}
\put(1.3,1.3){\sfr}
\end{picture}}}

\newcommand{\sthreetwobox}
{\stln \lower2.6ex\hbox{
\begin{picture}(3.6,2.6)
\multiput(.3,1.3)(1,0){3}{\sfr}
\put(.3,.3){\sfr}
\put(1.3,.3){\sfr}
\end{picture}}}

\newcommand{\sfourtwobox}
{\stln \lower2.6ex\hbox{
\begin{picture}(4.6,2.6)
\multiput(.3,1.3)(1,0){4}{\sfr}
\put(.3,.3){\sfr}
\put(1.3,.3){\sfr}
\end{picture}}}

\newcommand{\sfivetwobox}
{\stln \lower2.6ex\hbox{
\begin{picture}(5.6,2.6)
\multiput(.3,1.3)(1,0){5}{\sfr}
\put(.3,.3){\sfr}
\put(1.3,.3){\sfr}
\end{picture}}}

\newcommand{\ssixtwobox}
{\stln \lower2.6ex\hbox{
\begin{picture}(6.6,2.6)
\multiput(.3,1.3)(1,0){6}{\sfr}
\put(.3,.3){\sfr}
\put(1.3,.3){\sfr}
\end{picture}}}

\newcommand{\sthreethreebox}
{\stln \lower2.6ex\hbox{
\begin{picture}(3.6,2.6)
\multiput(.3,1.3)(1,0){3}{\sfr}
\multiput(.3,.3)(1,0){3}{\sfr}
\end{picture}}}

\newcommand{\sfourthreebox}
{\stln \lower2.6ex\hbox{
\begin{picture}(4.6,2.6)
\multiput(.3,1.3)(1,0){4}{\sfr}
\multiput(.3,.3)(1,0){3}{\sfr}
\end{picture}}}

\newcommand{\sfivethreebox}
{\stln \lower2.6ex\hbox{
\begin{picture}(5.6,2.6)
\multiput(.3,1.3)(1,0){5}{\sfr}
\multiput(.3,.3)(1,0){3}{\sfr}
\end{picture}}}

\newcommand{\ssixthreebox}
{\stln \lower2.6ex\hbox{
\begin{picture}(6.6,2.6)
\multiput(.3,1.3)(1,0){6}{\sfr}
\multiput(.3,.3)(1,0){3}{\sfr}
\end{picture}}}

\newcommand{\sfourfourbox}
{\stln \lower2.6ex\hbox{
\begin{picture}(4.6,2.6)
\multiput(.3,1.3)(1,0){4}{\sfr}
\multiput(.3,.3)(1,0){4}{\sfr}
\end{picture}}}

\newcommand{\sfivefourbox}
{\stln \lower2.6ex\hbox{
\begin{picture}(5.6,2.6)
\multiput(.3,1.3)(1,0){5}{\sfr}
\multiput(.3,.3)(1,0){4}{\sfr}
\end{picture}}}

\newcommand{\ssixfourbox}
{\stln \lower2.6ex\hbox{
\begin{picture}(6.6,2.6)
\multiput(.3,1.3)(1,0){6}{\sfr}
\multiput(.3,.3)(1,0){4}{\sfr}
\end{picture}}}

\newcommand{\sfivefivebox}
{\stln \lower2.6ex\hbox{
\begin{picture}(5.6,2.6)
\multiput(.3,1.3)(1,0){5}{\sfr}
\multiput(.3,.3)(1,0){5}{\sfr}
\end{picture}}}

\newcommand{\ssixfivebox}
{\stln \lower2.6ex\hbox{
\begin{picture}(6.6,2.6)
\multiput(.3,1.3)(1,0){6}{\sfr}
\multiput(.3,.3)(1,0){5}{\sfr}
\end{picture}}}

\newcommand{\ssixsixbox}
{\stln \lower2.6ex\hbox{
\begin{picture}(6.6,2.6)
\multiput(.3,1.3)(1,0){6}{\sfr}
\multiput(.3,.3)(1,0){6}{\sfr}
\end{picture}}}

\newcommand{\soneoneonebox}
{\stln \lower3.8ex\hbox{
\begin{picture}(1.6,3.6)
\multiput(.3,.3)(0,1){3}{\sfr}
\end{picture}}}

\newcommand{\stwooneonebox}
{\stln \lower3.8ex\hbox{
\begin{picture}(2.6,3.6)
\multiput(.3,.3)(0,1){3}{\sfr}
\put(1.3,2.3){\sfr}
\end{picture}}}

\newcommand{\stwotwoonebox}
{\stln \lower3.8ex\hbox{
\begin{picture}(2.6,3.6)
\multiput(.3,.3)(0,1){3}{\sfr}
\put(1.3,1.3){\sfr}
\put(1.3,2.3){\sfr}
\end{picture}}}

\newcommand{\stwotwotwobox}
{\stln \lower3.8ex\hbox{
\begin{picture}(2.6,3.6)
\multiput(.3,.3)(0,1){3}{\sfr}
\multiput(1.3,.3)(0,1){3}{\sfr}
\end{picture}}}

\newcommand{\sthreeoneonebox}
{\stln \lower3.8ex\hbox{
\begin{picture}(3.6,3.6)
\multiput(.3,.3)(0,1){3}{\sfr}
\put(1.3,2.3){\sfr}
\put(2.3,2.3){\sfr}
\end{picture}}}

\newcommand{\sthreetwoonebox}
{\stln \lower3.8ex\hbox{
\begin{picture}(3.6,3.6)
\multiput(.3,.3)(0,1){3}{\sfr}
\put(1.3,2.3){\sfr}
\put(2.3,2.3){\sfr}
\put(1.3,1.3){\sfr}
\end{picture}}}

\newcommand{\sthreetwotwobox}
{\stln \lower3.8ex\hbox{
\begin{picture}(3.6,3.6)
\multiput(.3,.3)(0,1){3}{\sfr}
\multiput(1.3,.3)(0,1){3}{\sfr}
\put(2.3,2.3){\sfr}
\end{picture}}}

\newcommand{\sthreethreeonebox}
{\stln \lower3.8ex\hbox{
\begin{picture}(3.6,3.6)
\multiput(.3,2.3)(1,0){3}{\sfr}
\multiput(.3,1.3)(1,0){3}{\sfr}
\put(.3,.3){\sfr}
\end{picture}}}

\newcommand{\sthreethreetwobox}
{\stln \lower3.8ex\hbox{
\begin{picture}(3.6,3.6)
\multiput(.3,2.3)(1,0){3}{\sfr}
\multiput(.3,1.3)(1,0){3}{\sfr}
\put(.3,.3){\sfr}
\put(1.3,.3){\sfr}
\end{picture}}}

\newcommand{\sthreethreethreebox}
{\stln \lower3.8ex\hbox{
\begin{picture}(3.6,3.6)
\multiput(.3,2.3)(1,0){3}{\sfr}
\multiput(.3,1.3)(1,0){3}{\sfr}
\multiput(.3,.3)(1,0){3}{\sfr}
\end{picture}}}

\newcommand{\sgencolbox}
{\stln \lower8.6ex\hbox{
\begin{picture}(1.6,7.6)
\multiput(.3,4.3)(0,1){3}{\sfr}
\put(.3,1.3){\framebox(1,3){$\vdots$}}
\put(.3,.3){\sfr}
\end{picture}}}

\newcommand{\sgenrowonebox}
{\stln \lower2.6ex\hbox{
\begin{picture}(7.6,2.6)
\multiput(.3,1.3)(1,0){3}{\sfr}
\put(3.3,1.3){\framebox(3,1){$\cdots$}}
\put(6.3,1.3){\sfr}
\put(.3,.3){\sfr}
\end{picture}}}

\newcommand{\sgenrowtwobox}
{\stln \lower2.6ex\hbox{
\begin{picture}(7.6,2.6)
\multiput(.3,1.3)(1,0){3}{\sfr}
\put(3.3,1.3){\framebox(3,1){$\cdots$}}

\put(6.3,1.3){\sfr}
\put(.3,.3){\sfr}
\put(1.3,.3){\sfr}
\end{picture}}}

\newcommand{\marcshapirowbox}
{\stln \lower2.6ex\hbox{
\begin{picture}(12.6,2.6)
\multiput(.3,1.3)(1,0){3}{\sfr}
\put(3.3,1.3){\framebox(3,1){$\cdots$}}
\put(6.3,1.3){\sfr}
\put(7.3,1.3){\sfr}
\put(8.3,1.3){\framebox(3,1){$\cdots$}}
\put(11.3,1.3){\sfr}
\put(7.3,1){$\underbrace{~~~~~~~~~~~~~~~}_{k}$}
\multiput(.3,.3)(1,0){3}{\sfr}
\put(3.3,.3){\framebox(3,1){$\cdots$}}
\put(6.3,.3){\sfr}
\put(.3,0){$\underbrace{~~~~~~~~~~~~~~~~~~~~~}_{n}$}
\end{picture}}}

\newcommand{\sgentworowonerowbox}
{\stln \lower2.6ex\hbox{
\begin{picture}(11.6,2.6)
\multiput(.3,1.3)(1,0){3}{\sfr}
\put(3.3,1.3){\framebox(3,1){$\cdots$}}
\put(6.3,1.3){\sfr}
\put(7.3,1.3){\framebox(3,1){$\cdots$}}
\put(10.3,1.3){\sfr}
\multiput(.3,.3)(1,0){3}{\sfr}
\put(3.3,.3){\framebox(3,1){$\cdots$}}
\put(6.3,.3){\sfr}
\end{picture}}}

\newcommand{\soneoneoneonebox}
{\stln \lower5ex\hbox{
\begin{picture}(1.6,4.6)
\multiput(.3,.3)(0,1){4}{\sfr}
\end{picture}}}

\newcommand{\soneoneoneoneoneonebox}
{\stln \lower7.4ex\hbox{
\begin{picture}(1.6,6.6)
\multiput(.3,.3)(0,1){6}{\sfr}
\end{picture}}}

\newcommand{\stwotwooneonebox}
{\stln \lower5ex\hbox{
\begin{picture}(2.6,4.6)
\multiput(.3,.3)(0,1){4}{\sfr}
\put(1.3,2.3){\sfr}
\put(1.3,3.3){\sfr}
\end{picture}}}

\newcommand{\sgenrowrowbox}
{\stln \lower2.6ex\hbox{
\begin{picture}(7.6,2.6)
\multiput(.3,1.3)(1,0){3}{\sfr}
\put(3.3,1.3){\framebox(3,1){$\cdots$}}
\put(6.3,1.3){\sfr}
\multiput(.3,.3)(1,0){3}{\sfr}
\put(3.3,.3){\framebox(3,1){$\cdots$}}
\put(6.3,.3){\sfr}
\end{picture}}}

\newcommand{\fourtwotwobox}
{\stln \lower3.8ex\hbox{
\begin{picture}(4.6,4.6)
\multiput(.3,.3)(0,1){4}{\fr}
\multiput(1.3,.3)(0,1){4}{\fr}
\put(2.3,3.3){\fr}
\put(3.3,3.3){\fr}
\end{picture}}}

\newcommand{\eq}{\begin{equation}}
\newcommand{\en}{\end{equation}}
\newcommand{\eqn}{\begin{eqnarray}}
\newcommand{\enn}{\end{eqnarray}}
\newcommand{\beq}{\begin{equation}}
\newcommand{\eeq}{\end{equation}}
\newcommand{\marcnbox}
{\stln \lower1.4ex \hbox{
\begin{picture} (6.6, 3.1)
\multiput(.3, .3) (1, 0) {2} {\fr}
\put(2.3, .3) {\framebox(3,1){$\cdots$}}
\put(5.3, .3){\fr}
\put(.3, 1.4)
{$\overbrace{~~~~~~~~~~~~~~~~~~}^{n}$}
\end{picture}}}
\newcommand{\marcnplusonebox}
{\stln \lower1.4ex \hbox{
\begin{picture} (6.6, 3.1)
\multiput(.3, .3) (1, 0) {2} {\fr}
\put(2.3, .3) {\framebox(3,1){$\cdots$}}
\put(5.3, .3){\fr}
\put(.3, 1.4)
{$\overbrace{~~~~~~~~~~~~~~~~~~}^{n+1}$}
\end{picture}}}
\newcommand{\marcnplustwobox}
{\stln \lower1.4ex \hbox{
\begin{picture} (6.6, 3.1)
\multiput(.3, .3) (1, 0) {2} {\fr}
\put(2.3, .3) {\framebox(3,1){$\cdots$}}
\put(5.3, .3){\fr}
\put(.3, 1.4)
{$\overbrace{~~~~~~~~~~~~~~~~~~}^{n+2}$}
\end{picture}}}
\newcommand{\marcnplusthreebox}
{\stln \lower1.4ex \hbox{
\begin{picture} (6.6, 3.1)
\multiput(.3, .3) (1, 0) {2} {\fr}
\put(2.3, .3) {\framebox(3,1){$\cdots$}}
\put(5.3, .3){\fr}
\put(.3, 1.4)
{$\overbrace{~~~~~~~~~~~~~~~~~~}^{n+3}$}
\end{picture}}}
\newcommand{\marcnplusfourbox}
{\stln \lower1.4ex \hbox{
\begin{picture} (6.6, 3.1)
\multiput(.3, .3) (1, 0) {2} {\fr}
\put(2.3, .3) {\framebox(3,1){$\cdots$}}
\put(5.3, .3){\fr}
\put(.3, 1.4)
{$\overbrace{~~~~~~~~~~~~~~~~~~}^{n+4}$}
\end{picture}}}
\newcommand{\smarctwojbox}
{\stln \lower1.4ex \hbox{
\begin{picture}(6.6,3.1)
\multiput(.3,.3)(1,0){2}{\sfr}
\put(2.3,.3){\framebox(3,1){$\cdots$}}
\put(5.3,.3){\sfr}
\put(.3,1.4){$\overbrace{~~~~~~~~~~~~~~~~~~}^{2j}$}
\end{picture}}}

\newcommand{\bea}{\begin{eqnarray}}
\newcommand{\eea}{\end{eqnarray}}
\newcommand{\ee}{\end{equation}}
\newcommand{\be}{\begin{equation}}

\newcommand{\no}{\nonumber}

\def\p{\partial}

\def\a{\alpha}
\def\b{\beta}

\def\no{\nonumber}

\def\ha{{\hat \alpha}}
\def\hb{{\hat \beta}}
\def\hg{{\hat \gamma}}
\def\hd{{\hat \delta}}

\def\da{{\dot \alpha}}
\def\db{{\dot \beta}}
\def\dg{{\dot \gamma}}
\def\dd{{\dot\delta}}

\begin{document}

\textwidth 170mm
\textheight 230mm
\topmargin -1cm
\oddsidemargin-0.8cm \evensidemargin -0.8cm
\topskip 9mm
\headsep9pt

\overfullrule=0pt
\parskip=2pt
\parindent=12pt
\headheight=0in \headsep=0in \topmargin=0in \oddsidemargin=0in

\vspace{ -3cm} \thispagestyle{empty} \vspace{-1cm}

\begin{flushright}
IGC-11/8-1 \ \ NSF-KITP-11-177
\end{flushright}

\begin{center}

{\Large \bf Superconformal symmetry  and maximal supergravity in various  dimensions}

\medskip

\vskip .2in

{\large  Marco Chiodaroli$^{1}$, Murat G\"{u}naydin$^2$ and Radu Roiban$^3$}
\let\thefootnote\relax\footnote{${}^1$email: mchiodar@gravity.psu.edu \hspace{0.15in} ${}^2$email: murat@phys.psu.edu \hspace{0.15in} 
${}^3$email: radu@phys.psu.edu }
\vskip .05in

{ \small Institute for Gravitation and the Cosmos \\
 The Pennsylvania State University \\
University Park PA 16802, USA }\\

\vskip .05in

and \\

\vskip .05in

{ \small Kavli Institute for Theoretical Physics \\
 University of California \\
 Santa Barbara CA 93106, USA }\\

\end{center}


\begin{abstract}

\vskip 0.05in
In this paper we explore the relation between conformal superalgebras with 64 supercharges
and maximal supergravity theories in three, four and six dimensions using twistorial oscillator techniques. 
The massless fields of ${\cal N}=8$ supergravity in four dimensions
were shown to fit  into a CPT-self-conjugate doubleton supermultiplet of the conformal superalgebra $SU(2,2|8)$ a long time ago.
We show that the fields of maximal supergravity in three dimensions can similarly 
be fitted into the super singleton multiplet of the conformal superalgebra $OSp(16|4,\mathbb{R})$, 
which is related to the doubleton supermultiplet of $SU(2,2|8)$ by dimensional reduction. 
Moreover, we construct  the ultra-short supermultiplet of the six-dimensional conformal superalgebra
$OSp(8^*|8)$ and show that its component fields can be organized in an on-shell superfield.
The ultra-short $OSp(8^*|8)$ multiplet reduces to the doubleton supermultiplet of $SU(2,2|8)$
upon dimensional reduction. We discuss the possibility  of a 
novel non-metric based $(4,0)$ six-dimensional supergravity theory with $USp(8)$ $R$-symmetry 
that reduces to maximal supergravity in four dimensions and is different from six-dimensional metric based
$(2,2)$ maximal supergravity, whose fields cannot be fitted into a unitary supermultiplet of a simple conformal superalgebra.
 
Such an interacting $(4,0)$ theory would be the non-metric gravitational analog of the interacting $(2,0)$ theory.

\end{abstract}

\baselineskip=16pt
\setcounter{equation}{0}
\setcounter{footnote}{0}

\newpage

\renewcommand{\theequation}{1.\arabic{equation}}
\setcounter{equation}{0}
\section{Introduction}

Superconformal invariance and the constraints it imposes on quantum field theories have
long been subject of active investigation.
In two dimensions, this symmetry
has been essential in the development of superstring theory.
In more recent years, the invariance of four-dimensional ${\cal N}=4$ super Yang-Mills theory
under the conformal superalgebra $PSU(2,2|4)$\footnote{By abuse of notation, 
we will use the same notation for supergroups and corresponding superalgebras throughout the paper.} has played an essential role in
formulating and testing the
$AdS/CFT$ correspondence \cite{mald,Witten:1998qj}. From 
the symmetry point of view, the $AdS/CFT$ correspondence 
is based on the isomorphism between conformal superalgebras in $d$ spacetime dimensions and 
$AdS$ superalgebras in $d+1$ dimensions. 

The study of Kaluza-Klein supergravity theories \cite{GM1,
gnw,mgnw} has long established the critical importance of some fundamental massless 
conformal supermultiplets in obtaining the spectra of 
type $IIB$ supergravity on $AdS_5\times S^5$ and eleven-dimensional supergravity on $AdS_{7(4)}\times S^{4(7)}$.  
The spectrum of $IIB$ supergravity on $AdS_5\times S^5$ was obtained by tensoring the CPT-self-conjugate 
doubleton supermultiplet of the four-dimensional,  
${\cal N}=4$ superconformal algebra $PSU(2,2|4)$ in \cite{GM1}, 
where it was pointed out that the doubleton supermultiplet does not have a Poincar\'e limit as a representation of the 
five-dimensional, ${\cal N}=8$ superalgebra. The authors of \cite{GM1} thus  suggested that the field theory of the doubleton supermultiplet lives on the boundary 
of $AdS_5$ which can be identified with Minkowski space and that the corresponding interacting theory is 
the ${\cal N}= 4$ super Yang-Mills theory which was known to be conformally invariant. 
Similarly, the spectra of eleven-dimensional supergravity over $AdS_7\times S^4$ and $AdS_4\times S^7$ were obtained by tensoring of doubleton and singleton supermultiplets of $OSp(8^*|4)$ and $OSp(8|4,\mathbb{R})$ \cite{gnw,mgnw}. 
These  supermultiplets do not have a Poincar\'e limit as $AdS_7$ and $AdS_4$ supermultiplets and 
their field theories were conjectured to be conformally invariant field theories on the boundaries which are the six- and three-dimensional Minkowski spaces\footnote{ The singular nature of 
the Poincar\'e limit of singleton (remarkable) representations of the four-dimensional $AdS$ group discovered by Dirac 
\cite{Dirac} was shown by Fronsdal and collaborators who pointed out that  their field theories
 live on the boundary of $AdS_4$ \cite{Fronsdal}.}.  For $OSp(8^*|4)$ with the even subgroup $SO(6,2) \times USp(4)$ 
the doubleton supermultiplet is the six-dimensional, $(2,0)$ conformal supermultiplet 
whose interacting theory is believed to be dual to $M$-theory on $AdS_7\times S^4$.
 Recently it was shown that the four-dimensional, ${\cal N}=4$ Yang-Mills supermultiplet and the six-dimensional, $(2,0)$ conformal supermultiplet are 
simply the minimal unitary  supermultiplets of 
$PSU(2,2|4)$ and $OSp(8^*|4)$, respectively \cite{FG2010,FG2010b}\footnote{A minimal unitary representation of a non-compact group is defined as a representation on the Hilbert 
space of functions of  the minimal number of variables possible. The minimal unitary supermultiplets of  non compact 
supergroups as defined and constructed in \cite{FG2010,FG2010b} contain  the minimal unitary representations and some of
 their deformations that are related by supersymmetry.}, which implies that the 
singleton and doubleton supermultiplets are of fundamental importance, not only physically but also mathematically.

The classification of potential counterterms in ${\cal N}=8$  
supergravity offers a novel
application of the four-dimensional superconformal algebra.
Based on its no-triangle behavior \cite{OneloopMHVGravity, NoTriangle, NoTriangleSixPt,  
BjerrumBohr:2008vc, Kallosh:2007ym},  it was suggested in \cite{BDR} that ${\cal N}=8$ 
supergravity may be finite
to all loop orders; 
indeed, subsequent explicit calculations show that its ultraviolet behavior matches
\cite{gravity3, gravity3m, gravity4, gravitysum} that of ${\cal N}=4$
super Yang-Mills theory up to at least four loops.
String dualities have also been used to argue for ultraviolet  
finiteness of ${\cal N}=8$
supergravity \cite{DualityArguments}, though difficulties with  
decoupling towers of
massive states may alter this conclusion \cite{GSO}.
Conservation of the Noether-Gaillard-Zumino duality current suggests
\cite{Kallosh:2011qt, Kallosh:2011dp} another argument for finiteness;  
a construction
of duality-invariant actions, potentially circumventing this argument,  
was outlined in \cite{Bossard:2011ij}.
The fields of ${\cal N}=8$ Poincar\'e supergravity can be fitted into  
a CPT-self-conjugate
doubleton supermultiplet of the conformal superalgebra $SU(2,2|8)$  
with $R$-symmetry
group $U(8)$ \cite{GM2,Drummond:2010fp}.
This superalgebra proves itself a valuable tool to study candidate  
counterterms
which would appear at the loop order at which the theory has its first  
divergence.
Potential on-shell counterterms should be local and should respect the  
on-shell symmetries
of the theory.
The authors  of \cite{Beisert:2010jx} have exhaustively analyzed all  
possible candidate counterterms
identifying local supersymmetry invariants with the top components of   
$SU(2,2|8)$
supermultiplets that are also $SU(8)$ singlets  and Lorentz scalars.
Their analysis shows that any counterterm below seven loops is incompatible
with the spontaneously broken $E_{7(7)}$ symmetry of the theory and  
recovers early results obtained
though a direct construction of $E_{7(7)}$ invariants respecting  
linearized supersymmetry~\cite{Brink:1979nt, Kallosh:1980fi}\footnote{An  
explicit potential 1/8-BPS seven-loop counterterm was recently  
constructed in \cite{Bossard:2011tq}. Potential eight-loop  
counterterms invariant under both $E_{7(7)}$ and supersymmetry have  
been constructed in \cite{Kallosh:1980fi, Howe:1980th}. Explicit  
proposals for potential linearized counterterms at three  
\cite{Kallosh:1980fi}, five and six loops \cite{Drummond:2003ex} have  
been ruled out based on their lack of $E_{7(7)}$ invariance in  
\cite{Brodel:2009hu, Elvang:2010kc} and \cite{Bossard:2010bd,  
Beisert:2010jx}, respectively.
The seven-loop bound on the order at which
the first divergence may appear has also been obtained through  
string-based amplitude computations
\cite{Green:2010wi, Green:2010sp} and through light-cone superspace  
arguments \cite{Kallosh:2009db}.
Together with $E_{7(7)}$ symmetry, lightcone superspace also appears  
to offer an argument for finiteness of ${\cal N}=8$ supergravity  
\cite{Kallosh:2009db}.}.

In this paper we will show that the fields of maximal supergravity in three dimensions also fit into a 
unitary supermultiplet of the conformal superalgebra $OSp(16|4,\mathbb{R})$. 
However, a naive extension of the analysis  of \cite{Beisert:2010jx} to dimensions higher than four 
is hindered by the fact that,
in dimensions higher than four, the maximal Poincar\'{e} supergravity multiplet is, in general,  not a representation
of the maximal superconformal algebra that has the full $R$-symmetry group as a factor in its even subgroup.

Interestingly, a similar issue arises considering ${\cal N}=4$ super Yang-Mills theory in
four dimensions: while maximally supersymmetric Yang-Mills theories exist in dimensions
other than four, their supermultiplets are not representations of the corresponding
superconformal algebras.
In six dimensions, the theory which is constrained by
$OSp(8^*|4)$ and is the analog of four-dimensional super Yang-Mills theory is the the mysterious interacting
$(2,0)$ theory discovered in \cite{Witten:1995zh,Strominger:1995ac}.
Both the conformal $(2,0)$ theory and the non-conformal six-dimensional super Yang-Mills theory yield
the conformal ${\cal N}=4$ super Yang-Mills theory when reduced to four dimensions.

In this paper, we argue that three- and four-dimensional, maximal  Poincar\'e supergravity 
theories may have a similar six-dimensional
counterpart whose fields at the linearized level belong to a unitary supermultiplet of six-dimensional superconformal group  $OSp(8^*|8)$. To this end, we  construct the unitary irreducible representation of 
$OSp(8^*|8)$ which involves only fields of spin less or equal to two and which
reduces to the CPT-self-conjugate doubleton supermultiplet of $SU(2,2|8)$ in four dimensions. 
This supermultiplet is equivalent to the  supermultiplet  of $(4,0)$ linearized conformal supergravity 
studied earlier by Hull using the formalism of dual gravitons \cite{Hull:2000zn,Hull:2000rr}\footnote{We thank Henning Samtleben for bringing this work to our attention.}.
This multiplet may be thought of as the direct product of two $(2,0)$  multiplets, in the same
sense as the four-dimensional supergravity multiplet can be interpreted as the direct product
of two ${\cal N}=4$ Yang-Mills multiplets.
The existence of this six-dimensional supermultiplet suggests the tantalizing possibility that there exists a novel  interacting
six-dimensional supergravity theory with maximal supersymmetry and $USp(8)$ $R$-symmetry,
whose field content becomes, upon dimensional reduction, that of ${\cal N}=8$ supergravity in
four dimensions, just as the fields of the six-dimensional, $(2,0)$ theory reduce to those 
of ${\cal N}=4$ super Yang-Mills in four dimensions. 
However, if such a  $(4,0)$ interacting theory exists, it would not be a metric-based gravitational theory in the usual sense.
This follows from the fact that, just like the six-dimensional Yang-Mills theory, ordinary Einstein gravity based on a metric is not conformally invariant even at the linearized level\footnote{ Even though the interacting $(4,0)$ theory is not metric based, we shall, by an abuse of terminology, refer to it as $(4,0)$ supergravity.}. Just like the interacting $(2,0)$ theory we do not expect this novel interacting $(4,0)$ theory to admit a covariant action.
We should stress that our main starting point,  in all dimensions, 
is the construction of the relevant unitary representations  of the  conformal and superconformal groups.
We show explicitly  how these unitary representations are realized as covariant conformal fields and  conformal 
supermultiplets of fields by going to the coherent state bases of the corresponding conformal groups and supergroups. 
%
Our formalism constructs directly the gauge-invariant field strengths 
associated with the fields of the relevant supergravity multiplets 
and hence guarantees that only the physical degrees of freedom appear in the representations.

In Section \ref{4dconformalgroup}, we review the oscillator construction of the unitary representations of the four-dimensional conformal
 group $SU(2,2)$. In this approach, unitary irreducible representations (UIRs) of the conformal group are constructed using a
set of twistorial bosonic oscillators which obey canonical commutations relations.
The conformal fields correspond to coherent states labelled by the coordinate four-vector obtained  by  acting with the translation generator $e^{-i x_\mu P^\mu}$ on a set of states $|\Phi^{\ell}_{j_M,j_N}(0) \rangle $ transforming covariantly  in the $(j_M,j_N)$ representation of  the Lorentz group $SL(2,\mathbb{C})$ with a definite scale dimension $\ell$. The states $|\Phi^{\ell}_{j_M,j_N}(0) \rangle $ can be written simply in the form $T|\Omega (j_L,j_R,E )\rangle $ where $|\Omega (j_L,j_R,E )\rangle $ are a set of states transforming irreducibly under the maximal compact subgroup $SU(2)_L\times SU(2)_R \times U(1)_E$ of  $SU(2,2)$ such that the labels $(j_L,j_R,E)$ coincide with labels $(j_M,j_N,-\ell)$ and $T$ is the intertwining operator between compact and manifestly unitary basis and the non-compact and manifestly covariant basis. The review of the oscillator construction in the compact basis is relegated to Appendix A.
In Section \ref{multipletN8}, we discuss  the extension of oscillator method   to the conformal
superalgebra $SU(2,2|8)$  and  review the results of \cite{GM2} on the construction of the CPT-self-conjugate doubleton supermultiplet of $SU(2,2|8)$. The fields of ${\cal N}=8$ Poincar\'e supergravity
are in one-to-one correspondence with the fields of this CPT-self-conjugate doubleton supermultiplet. This supermultiplet has recently been used in the construction and classification of counterterms in ${\cal N}=8$ supergravity \cite{Beisert:2010jx}.

Section \ref{3dconformalgroup} is devoted to the construction of the unitary supermultiplets of the three-dimensional superconformal group  $OSp(16|4,\mathbb{R})$ with even subgroup 
$SO(16)\times SO(3,2)$. We show that the fields of maximal supergravity can be identified with the supersingleton multiplet and the four-dimensional doubleton supermultiplet of $SU(2,2|8)$ reduces   to this singleton supermultiplet under dimensional reduction.

In Section \ref{6dconformalgroup}, we review the  unitary irreducible representations of the six-dimensional conformal
superalgebras  $OSp(8^*|2N)$ and construct the  $(4,0)$ "chiral graviton supermultiplet"  of $OSp(8^*|8)$ with the $R$-symmetry group $USp(8)$. 
In Section \ref{6dsuperfield},
using a six-dimensional superspace,  we show how the fields of the $(4,0)$ supermultiplet can be organized into an on-shell
superfield obeying one algebraic and one differential constraint.
In Section \ref{secpot} we write down the gauge potentials corresponding to the fields of the $(4,0)$ supermultiplet 
and show that they can be written as fields with mixed-symmetry Lorentz indices. 
We then study the dimensional reduction of six-dimensional, $(4,0)$ supermultiplet down to four dimensions. 
We  show that the $(4,0)$ supermultiplet reduces to the CPT-self-conjugate doubleton supermultiplet of $SU(2,2|8)$.

In Section \ref{interacting} we comment on the possibility of an interacting theory 
of the $(4,0)$ supermultiplet and on the restrictions on the allowed interactions from 
the higher-spin no-go theorems in the literature. We then discuss the implications of our results.

\renewcommand{\theequation}{2.\arabic{equation}}
\setcounter{equation}{0}

\bigskip 

\section{Unitary representations of $SU(2,2|8)$ and ${\cal N}=8$
 supergravity in four dimensions  \label{4dconformalgroup}}

\subsection{
$(SL(2,\mathbb{C}) \times \mathcal{D})$ covariant coherent states of $SU(2,2)$ and $4D$  conformal fields}

The generators of the conformal group in four dimensions $SU(2,2)$ (the two sheeted
covering of $SO(4,2)$) satisfy the commutation relations
\eq
[M_{ab}, M_{cd}] = i(\eta_{bc}M_{ad} - \eta_{ac}M_{bd}
-\eta_{bd}M_{ac} + \eta_{ad}M_{bc}) \ .\label{SO42}
\en
We shall use the Minkowski metric $\eta_{a b}=\textrm{diag}(+,-,-,-,-,+)$ with $a,b=0,1,2,3,5,6$. 
$M_{\mu\nu}$, with $\mu, \nu, \dots$ $= 0, 1, 2, 3$, are the Lorentz group generators.
The four-momentum generators $P_{\mu}$, the special conformal generators $K_{\mu}$ 
and the dilatation generator $D$ are given by,
\eq
M_{\mu 5} = {1 \over 2} (P_{\mu} - K_{\mu})\ , \quad
M_{\mu 6} = {1 \over 2} (P_{\mu} + K_{\mu})\ , \quad M_{56} = -D \ .
\en
The fields of a four-dimensional  conformal field theory should  transform covariantly
under the Lorentz group $SL(2,\mathbb{C})$ and dilatations \cite{macksalam}.
The corresponding unitary realizations of the conformal group are
induced from finite dimensional representations of the stability group
$\mathcal{H}$ which is the semi-direct product of Lorentz group and
dilatations  with the Abelian group $\mathcal{K}_{4}$ of special conformal
transformations. In other words,
conformal  fields live on the
coset space $G/\mathcal{H}$, which in our case is just the conformal compactification
of four-dimensional Minkowski spacetime. Consequently, these representations are labelled
by their $SL(2,\mathbb{C})$ labels $(j_{M},j_{N})$, their conformal
dimension $l$ and certain matrices $\kappa_{\mu}$ related to their behavior
under special conformal transformations generated by $K_{\mu}$.

The stability group $\mathcal{H}$ of the origin $x^{\mu}=0$ under the action of the conformal group  $SO(4,2)$ is the semi-direct product group  
\[
\mathcal{H} = \left(SL(2,\mathbb{C}) \times \mathcal{D}\right) \circledS  \mathcal{K}_4~~;
\]
the Lie algebra of the stability group $\mathcal{H}$  is simply the semi-direct sum
 of the generators $ M_{\mu\nu}$ of the Lorentz group $SL(2,\mathbb{C})$, the  generator   $D$ of   dilatations  and 
the generators $K_\mu$ of  special conformal transformations.
We shall take  complex linear combinations of the generators of the Lorentz group $SL(2,\mathbb{C})$ to represent them  as $SU(2)_M\times SU(2)_N$
 generators
\eqn
M_{m}=\frac{1}{2} \left( \frac{1}{2} \varepsilon_{mnl} M_{nl}+ i M_{0m}
\right)
\ , \qquad
N_{m}=\frac{1}{2} \left( \frac{1}{2} \varepsilon_{mnl} M_{nl}- i M_{0m}
\right) \ .
\enn
They satisfy
\eqn [M_{m},M_{n}]=i \varepsilon_{mnl}M_{l} \ , \qquad   [N_{m},N_{n}]=i \varepsilon_{mnl}N_{l} \ , \qquad [M_{m},N_{n}]=0 \ .
\enn

In Appendix \ref{AppA} we review the oscillator construction of the positive energy unitary representations of the conformal group in a manifestly unitary compact $SU(2)_L\times SU(2)_R\times U(1)_E$ basis.
To relate the positive energy unitary representations constructed in the manifestly unitary compact basis  to conformal fields transforming covariantly with respect to the Lorentz group $SL(2,\mathbb{C})$ with a definite scale dimension given by the eigenvalues of the dilatation generator $D$ we use the operator \cite{GMZ2}
\eq
T =e^{\frac{\pi}{4}M_{05}} \ ,
\label{4dintertwiner}
\en
which corresponds to a purely imaginary rotation through ${\pi \over 4}$.
It satisfies
\begin{eqnarray}
M_{m} T  =   T L_{m} \quad,\quad
N_{m} T  =   T R_{m} \quad,\quad
D T        =   T E \label{UL} \ ,
\end{eqnarray}
where $L_m$, $R_m$  and $E$ are the compact $SU(2)_L\times SU(2)_R \times U(1)_E$ generators.
In other words $T$ "intertwines" the representations of the Lorentz group and dilatation generators
with those of the maximal compact subgroup
$SU(2)_L\times SU(2)_R \times U(1)_E $. 
Furthermore $T$ intertwines the momentum generators $P_\mu$ and the special conformal generators $K_\mu$ 
into operators belonging respectively to the grade $+1$ and the grade $-1$  
subspaces $\mathcal{L}^\pm$ in the compact basis given in Appendix \ref{AppA}. We obtain the expressions,
\be
\big( \sigma_\mu P^\mu \big)^{\a \b} T = T L^{i,j} \ , \qquad 
\big( \bar \sigma_\mu K^\mu \big)_{\a \b} T = T L_{i,j} \ . \label{UL2}
\ee
The generators are realized in terms of oscillators obeying canonical commutation relations.
The operator $T$ intertwines the  oscillators that transform covariantly under the compact subgroup 
$SU(2)_L \times SU(2)_R $ with the oscillators that transform covariantly under the Lorentz group $SL(2,\mathbb{C})$. 
The oscillators $a_i (a^i)$ and $b_i (b^i)$ that transform in the $({1 \over 2},0)$ and $(0,{1 \over 2}) $ 
representation of $SU(2)_L\times SU(2)_R$  go over to covariant oscillators transforming as Weyl spinors $({1 \over 2},0)$ and $(0,{1 \over 2})$ of the Lorentz group, respectively. Hence we shall label the covariant indices of Weyl spinors with undotted and dotted 
Greek indices $\alpha, \beta, \dots \dot{\alpha}, \dot{\beta}, \dots $,
\eqn
\eta_\alpha &= &T a_i T^{-1} = \frac{1}{\sqrt{2}} ( a_i - b^i ) \ , \nn \\
\lambda^{\alpha} &= &T a^i T^{-1} = \frac{1}{\sqrt{2}} ( a^i + b_i ) \nn \ , \\
\tilde{\eta}_{\dot{\alpha}}&=& T b_i T^{-1} = \frac{1}{\sqrt{2}} ( b_i - a^i ) \ , \\
\tilde{\lambda}^{\dot{\alpha}}&=& T b^i T^{-1} = \frac{1}{\sqrt{2}} ( b^i + a_i ) \ ; \nn
\enn
where $i,j,\dots ;\alpha, \beta, \dots ; \dot{\alpha}, \dot{\beta}, \dots =1,2$ and the covariant indices on
the left hand side match the indices $i,j$ on the right hand side of the equations\footnote{Here and elsewhere in this paper $\lambda$ and ${\tilde\lambda}$ will
denote operators whose eigenvalues are the familiar bosonic spinors parameterizing
momenta in  the spinor helicity formalism. The operators $\eta$ are however unrelated
to the Grassmann-odd parameters appearing in the supersymmetric extension of this formalism.}.
They satisfy the covariant commutation relations
\eqn
[ \eta_\alpha , \lambda^\beta ]  =  \delta^\beta_\alpha \ , \qquad  \quad
{[} \tilde{\eta}_{\dot{\alpha}}, \tilde{\lambda}^{\dot{\beta}} {]} = 
\delta_{\dot{\alpha}}^{\dot{\beta}}~~.
\enn
Furthermore we have
\eqn
( \sigma^\mu P_\mu ) ^{\alpha \dot{\beta}}
= \lambda^\alpha \tilde{\lambda}^{\dot{\beta}}  \ , \qquad \quad 
( \bar{\sigma} ^\mu K_\mu ) _{\alpha \dot{\beta}}
= \eta_\alpha \tilde{\eta}_{\dot{\beta}} \ .
\enn
Acting with $T$ on a lowest energy irreducible representation  $|\Omega (j_L,j_R,E )\rangle$ in the compact basis corresponds
to a pure imaginary rotation in the corresponding representation space of $SU(2,2)$,
\[
T|\Omega (j_L,j_R,E )\rangle = e^{ \frac{\pi}{4} M_{05} } |\Omega (j_L,j_R,E )\rangle \ .
\]
Since the negative grade generators in the compact basis annihilate the lowest energy irreducible representation,
\[\mathcal{L}^{-}|\Omega (j_L,j_R,E )\rangle =0 \ , \]
 it then follows from (\ref{UL}) that $| \Phi^\ell_{j_M,j_N}(0) \rangle :=
T|\Omega (j_L,j_R,E )\rangle$ is an
irreducible representation of the little group $\mathcal{H}$ with $SL(2,\mathbb{C})$
quantum numbers $(j_{M},j_{N})=(j_{L},j_{R})$, conformal dimension\footnote{
In our conventions, $\ell$ is the length (or inverse mass) dimension.}  $\ell=-E$
and trivially represented special conformal transformations $K_{\mu}$
({\it i.e.} $\kappa_{\mu}=0$) as a consequence of equation~(\ref{UL}),
\eq
K_\mu |\Phi^\ell_{j_M,j_N}(0) \rangle =0~~.
\en
Acting with $e^{-i x^{\mu}P_{\mu}}$ on $|\Phi^\ell_{j_M,j_N}(0) \rangle $ one forms a coherent state labelled by the coordinate $x_\mu$, which we will denote as $|\Phi^\ell_{j_M,j_N}(x_\mu)  \rangle$,
\eq
e^{-i x^{\mu}P_{\mu}}|\Phi^\ell_{j_M,j_N}(0) \rangle \equiv |\Phi^{\ell}_{j_M,j_N}(x_\mu) \rangle \ .
\en
 These coherent states transform exactly like the states created by the action of  
conformal fields $\Phi^{\ell}_{j_M,j_N}(x_\mu) $ acting on the vacuum vector $| 0\rangle $,
 \[ \Phi^{\ell}_{j_M,j_N}(x^{\mu}) | 0 \rangle  \cong |\Phi^{\ell}_{j_M,j_N}(x_\mu) \rangle \ , \]
with exact numerical coincidence of the compact
and the covariant labels $(j_{L},j_{R},E)$ and $(j_{M},j_{N},-l)$, respectively \cite{GMZ2}.

Let us now  derive the explicit transformation properties of the coherent states under all the generators of the conformal group. First from the commutation relations of $SU(2,2)$ generators we have
\eq
e^{i x \cdot P} K_\mu e^{-i x \cdot P} = K_\mu - 2 x_\mu D + 2 x^\nu M_{\nu \mu} + 2 x_\mu (x \cdot P) - x^2 P_\mu \ ,
\en
where $x \cdot P = x_\mu P^\mu$. This implies that
\eqn
K_\mu|\Phi^{\ell}_{j_M,j_N}(x_\mu) \rangle
 & = & e^{-i x\cdot P} \left( K_\mu - 2 x_\mu D + 2 x^\nu M_{\nu \mu} + 2 x_\mu (x \cdot P) - x^2 P_\mu \right) T|\Omega (j_L,j_R,E )\rangle \nn \\
& = &  \left( - x^2 P_\mu + 2 x_\mu ( x\cdot P) + 2 x_\mu \ell + 2 x^\nu \Sigma_{\nu\mu} \right) |\Phi^{\ell}_{j_M,j_N}(x_\mu) \rangle
\ , \enn
where $\Sigma$ represents the $SL(2,\mathbb{C})$ matrices in the representation
$ (j_M,j_N) = ( j_L,j_R)$ and $\ell$ is the conformal dimension. The last equation above can be recast in the form,
\eq
K_\mu |\Phi^{\ell}_{j_M,j_N}(x_\mu) \rangle
 = \left( -i\frac{\partial}{\partial x^\mu} + 2 i x_\mu x^\nu \frac{\partial}{\partial x^\nu}
 + 2 \ell x_\mu + 2 x^\nu \Sigma_{\nu\mu} \right) |\Phi^{\ell}_{j_M,j_N}(x_\mu) \rangle~~.
\en
Similarly we have
\eq
P_\mu |\Phi^{\ell}_{j_M,j_N}(x_\mu) \rangle
 = i \frac{\partial}{\partial x^\mu} |\Phi^{\ell}_{j_M,j_N}(x_\mu) \rangle \ .
\en
Using
\eq
D T = T E
\en
and
\eq
e^{i x\cdot P} D e^{-i x\cdot P} = D - x^\mu P_\mu
\en
one finds
\eq
D |\Phi^{\ell}_{j_M,j_N}(x_\mu) \rangle
 = - \left( i x^\mu \frac{\partial}{\partial x^\mu} + \ell \right) |\Phi^{\ell}_{j_M,j_N}(x_\mu) \rangle~~.
\en
Using
\eq
e^{i x\cdot P } M_{\mu\nu} e^{-i x \cdot P} = M_{\mu\nu} - x_\mu P_\nu + x_\nu P_\mu
\en
one finds
\eq
M_{\mu\nu} |\Phi^{\ell}_{j_M,j_N}(x_\mu) \rangle
= \left( \Sigma_{\mu\nu} - i x_\mu \frac{\partial}{\partial x^\nu} + i x_\nu \frac{\partial}{\partial x^\mu} \right) |\Phi^{\ell}_{j_M,j_N}(x_\mu) \rangle~~.
\en

  The lowest weight UIRs
of $SU(2,2)$ with vanishing four-dimensional Poincar\'e mass $m^{2}=P_{\mu}P^{\mu}$
are exactly
the ones obtained by taking only one generation (or color) of oscillators ($a_i $ and $b_j$)
\cite{GM1,GMZ2,Binegar:1981gv}.
These are precisely the doubleton representations \cite{GM1,GMZ2,GMZ1}. They  were
studied in the 1960s as "ladder representations" of the conformal group \cite{barut,macktodorov}.
More recently, it was shown that the doubleton unitary representation correspond to
the minimal unitary representation of $SU(2,2)$ and one parameter deformations
thereof \cite{FG2010}. The deformation parameter is simply the helicity of the
corresponding massless conformal field.


\subsection{CPT-self-conjugate doubleton supermultiplet of $SU(2,2|8)$ and fields of $4D$, ${\cal N}=8$
supergravity \label{multipletN8}}

In this section we shall review the results of \cite{GM2} where it was shown
that the massless fields of ${\cal N}=8$ four-dimensional supergravity can be organized
into the CPT-self-conjugate doubleton supermultiplet of the ${\cal N}=8$ superconformal algebra
$SU(2,2|8)$.
It is  not clear whether this fact has direct implications for the interacting theory of this multiplet.
Indeed, the corresponding supermultiplet of $PSU(2,2|4)$ is the ${\cal N}=4$ Yang-Mills
supermultiplet \cite{GM1}; the interacting theory of the CPT-self-conjugate doubleton
supermultiplet of $PSU(2,2|4)$ is known to be conformally invariant.
However, the interacting non-linear maximal ${\cal N}=8$ supergravity is not conformally
invariant.\footnote{An interesting possibility is that the violation if scale invariance occurs only 
at the classical level; a consequence of such a scenario would be that the theory is finite.}

The construction of representations of the $SU(2,2|8)$ superalgebra makes use of its
three graded decomposition with respect to its maximal compact subsuperalgebra
$\mathcal{L}^{0} =SU(2|p)\times SU(2|8-p) \times U(1)$
\eq
SU(2,2|8) = \mathcal{L}^{+} \oplus \mathcal{L}^{0} \oplus \mathcal{L}^{-}~~,
\en
where
\eqn
[\mathcal{L}^{0},\mathcal{L}^{\pm}]=
\mathcal{L}^{\pm} \ ,
\qquad\qquad
[\mathcal{L}^{+},\mathcal{L}^{-}]=
\mathcal{L}^{0} \ ,
\qquad\qquad
[\mathcal{L}^{+},\mathcal{L}^{+}]=
0=[\mathcal{L}^{-},\mathcal{L}^{-}]~~.
\enn

The Lie superalgebra $SU(2,2|8)$
 can be realized in terms of bilinear combinations of bosonic and
fermionic annihilation and creation operators (superoscillators) $\xi_{A}$
($\xi^{A}={\xi_{A}}^{\dagger}$) and $\eta_{M}$
($\eta^{M}={\eta_{M}}^{\dagger}$)
which transform covariantly (contravariantly)
under the  $SU(2|p)$ and $SU(2|8-p)$ sub-supergroups of $SU(2,2|8)$. We denote the annihilation and creation operators with lower and
upper indices, respectively. 
Each superoscillator may be represented as a doublet as
\eq
\xi_{A} =  \left(\begin{array}{c}a_{i} \\ \alpha_{\hat{x}}\end{array}\right) \ ,
\quad
\xi^{A} = \left(\begin{array}{c}a^{i} \\
                        \alpha^{\hat{x}}\end{array} \right)
\en
and
\eq
\eta_{M} = \left(\begin{array}{c}b_{i} \\
                        \beta_{x}  \end{array}  \right) \ , \quad
\eta^{M} = \left(\begin{array}{c}b^{i} \\
                        \beta^{x} \end{array} \right) \ ,
\en
with $i,j=1,2$; $\hat{x},\hat{y}=1,2,..,p$; $x,y=1,2,..,8-p$ and
\be [a_i, a^j] = \delta_{i}^{j} \ , \quad
\{\alpha_{\hat{x}}, \alpha^{\hat{y}}\} = \delta_{\hat{x}}^{\hat{y}} \ ,  \nn \ee
\be  [b_i, b^j] = \delta_{i}^{j} \ , \quad 
 \{\beta_{x}, \beta^{y}\} = \delta_{x}^{y} \ .  \ee
The generators of $SU(2,2|8)$ are given in terms of the above
superoscillators
as
\eqn
\mathcal{L}^{-} = \xi_{A} \eta_{M}
\quad,\quad\qquad
\mathcal{L}^{0} = \xi^{A} \xi_{B}
\oplus \eta^{M} \eta_{N}
\quad,\quad\qquad
\mathcal{L}^{+} = \xi^{A} \eta^{M}~~.
\enn

The $SU(8)$ generators, written in terms of fermionic
oscillators $\alpha$ and $\beta$, read as follows,
\eqn
A^{\hat{y}}_{\hat{x}} &=& \alpha^{\hat{y}} \alpha_{\hat{x}}
-{1 \over p} \delta^{\hat{y}}_{\hat{x}} N_{\alpha} \ , \cr
B^{y}_{x} &=& \beta^{y} \beta_{x}
-{1 \over q} \delta^{y}_{x} N_{\beta} \ ,  \cr
C &=& -{1 \over p} N_{\alpha} - {1 \over (8-p) } N_{\beta} +1 \ , \cr
L_{\hat{x} x} &=& \alpha_{\hat{x}} ~ \beta_{x} \ , \cr 
L^{x \hat{x}} &=& \beta^{x} ~\alpha^{\hat{x}} \ ,
\enn
where $N_{\alpha}=\alpha^{\hat{y}} \alpha_{\hat{y}} $
and $N_{\beta} = \beta^{x} \beta_{x}$
are the fermionic number operators.

Similarly, the $SU(2,2)$ generators, written in terms of
bosonic oscillators $a$ and $b$, read
\eqn
L_{ij} &=& a_{i}  b_{j} \ , \cr
L^{ij} &=& a^{i} b^{j}  \ ,  \cr 
L^{i}_{j} &=& a^{i}  a_{j} 
-{1 \over m} \delta^{j}_{i} N_{a} \ , \cr
R^{i}_{j} &=& b^{i} b_{j} 
-{1 \over n} \delta^{i}_{j} N_{b} \ , \cr
E &=& {1 \over 2} N_a + {1 \over 2} N_b + 1 \ ,
 \enn
where $N_{a} \equiv a^{i}  a_{i},
N_{b} \equiv b^{i}  b_{i}$ are the
bosonic number operators.

Starting from  a set of vectors  $ |\Omega \rangle$
in the Fock space transforming irreducibly under $SU(2|4) \times
SU(2|4) \times U(1)$ and annihilated by $\mathcal{L}^{-}$, one can
generate the doubleton UIRs of $SU(2,2|8)$ by repeated application of
$\mathcal{L}^{+}$
\eq
 |\Omega \rangle ,\quad  \mathcal{L}^{+1}|\Omega \rangle ,\quad
\mathcal{L}^{+1} \mathcal{L}^{+1}|\Omega \rangle , ... \ .
\en
The irreducibility of doubleton UIRs of $SU(2,2|8)$ as constructed is  a consequence of the
irreducibility of $|\Omega \rangle$ under $SU(2|4) \times
SU(2|4) \times U(1)$.

In  the subspace involving purely bosonic
oscillators one gets the subalgebra $SU(2,2)$ and the above
construction yields its positive energy doubleton UIRs.
In  the subspace involving purely
fermionic oscillators one  gets the compact $R$-symmetry group $ SU(8)$ and
 the resulting  representations of $SU(8)$ in
its $SU(4)\times SU(4) \times U(1)$ basis.

The positive energy UIRs of $SU(2,2|8)$ decompose into a direct sum of
finitely many positive energy UIRs of $SU(2,2)$
transforming in definite irreducible representations
of the $R$-symmetry group $SU(8)$.
Thus, each positive energy UIR of $SU(2,2|8)$
corresponds to a supermultiplet of four-dimensional massless fields.

The CPT-self-conjugate supermultiplet of $SU(2,2|8)$ is obtained by choosing 
the Fock vacuum as the lowest weight vector  $|\Omega \rangle =|0\rangle$ in the 
$SU(2|4)\times SU(2|4) \times U(1)$ basis; it was given in  \cite{GM2} and we reproduce it in Table \ref{Table_GM}. \\

\begin{table}[ht]
\begin{center}
\begin{tabular}{|c|c|c|c|c|c|}
\hline
{Lowest weight vector}  &   ${ SU(2)_{j_1}\times SU(2)_{j_2} } $ & ${ E_0} $ & $ {  SU(8)}$ & ${ U(1)_A}$ &{ Fields}
\\ \hline
$|0\rangle $          & $(0,0)$ & 1     & ${\bf 70}$   & 0 &$\phi^{[ABCD]} $
\\ \hline
$L^{ix} |0\rangle$   & $({1 \over 2},0)$ &  ${3 \over 2} $& ${ \bf 56}$ & 1 &$\lambda^{[ABC]}_{+} \equiv \lambda_{\alpha}^{[ABC]}$
\\ \hline
$L^{i \hat{x}} |0\rangle$        & $(0,{1 \over 2})$   & ${3 \over 2}$   &$\overline{ \bf 56}$ & -1 &$\lambda_{-[ABC]} \equiv \lambda_{\dot{\alpha}[ABC]}$
\\ \hline
$ (L^{ix})^2 |0\rangle$ & (1,0)  & 2 & ${\bf 28}$& 2 &$h_{\mu\nu}^{+[AB]} \equiv h_{(\alpha\beta)}^{[AB]}$
\\ \hline
$ (L^{i \hat{x}})^2 |0\rangle$ & (0,1)  & 2 & $\overline{\bf 28}$& -2 &$h_{\mu\nu [AB]}^{-} \equiv h_{(\dot{\alpha}\dot{\beta})[AB]}$
\\ \hline
$ (L^{ix})^3 |0\rangle$ & $({3 \over 2},0)$  & ${5 \over 2}$ & ${\bf 8}$& 3 &$\partial_{[ \mu}\psi_{\nu]}^{+ A} \equiv \psi_{(\alpha\beta\gamma)}^{A}$
\\ \hline
$ (L^{i \hat{x}})^3 |0\rangle$ & $(0,{3 \over 2})$  & ${5 \over 2}$ & $\bar{\bf 8}$& -3 &$\partial_{[ \mu}\psi_{\nu] A}^{-} \equiv \psi_{(\dot{\alpha}\dot{\beta}\dot{\gamma}) A}$
\\ \hline
$ (L^{ix})^4 |0\rangle$ & $(2,0)$  &  3 & $1 $& 4 &$ R_{(\alpha\beta\gamma\delta)}$
\\ \hline
$ (L^{i \hat{x}})^4 |0\rangle$ & $(0,2)$  & 3 & $1 $& -4 &$ R_{(\dot{\alpha}\dot{\beta}\dot{\gamma}\dot{\delta} )}$
\\ \hline
\end{tabular}
\medskip
\caption{\small \label{Table_GM} 
The CPT-self-conjugate doubleton supermultiplet of $SU(2,2|8)$ defined by 
the lowest weight vector $ |0\rangle$ in the $SU(2|4)\times SU(2|4)\times U(1)$ basis.
The first column indicates the lowest weight vectors of
$SU(2,2) \times SU(8)$. The oscillator formalism gives directly the gauge-invariant fields 
strengths associated with the fields in the representation. $A,B,C,..=1,2,..,8 $ are the SU(8) $R$-symmetry  indices. The indices $\alpha, \beta,..$ and $\da, \db, ..$ denote the chiral and anti-chiral spinorial indices of $SL(2,\mathbb{C})$. Round (square) brackets indicate symmetrization (antisymmetrization) of  the enclosed indices.}
\end{center}
\end{table}
On-shell fields of linearized ${\cal N}=8$ supergravity in four dimensions satisfy massless free field equations that are conformally invariant. 
The nonlinear completion of the ${\cal N}=8$ supergravity breaks the conformal supersymmetry algebra $SU(2,2|8)$ to ${\cal N}=8$ Poincar\'e supersymmetry.
There is no known conformal supergravity with the same field content as maximal supergravity in four dimensions.
The superalgebra $SU(2,2|8)$ and the  supermultiplet of fields given in Table \ref{Table_GM}
have been used in the construction and analysis of potential higher-loop counterterms to maximal
supergravity \cite{Drummond:2010fp,Beisert:2010jx}.

\subsection{Four-dimensional constrained superfields}

It has long been known that fields of the CPT-self-conjugate doubleton supermultiplet of $SU(2,2|8)$ can be assembled into a scalar 
superfield $W_{abcd}$ \cite{Siegel:1980bp,Howe:1981qj,fs} which depends on the ${\cal N}=8$ superspace coordinates,
\be \big( x^{\a \dot \b}, \theta^{\a a} , \bar \theta^{\dot \b}_b   \big) \ee 
with $\a,\dot \b= 1,2 $ and $a=1,2,\dots 8$. One can introduce superspace covariant derivatives
obeying the relation
\be \{ D_{\a a}, \bar{D}_{\dot \b}^b \} = 2 i \delta_{a}^b \partial^{\vphantom{A}}_{\a \dot \b} \ .
\label{comm-D-4d}  \ee
The superfield $W_{abcd}$ transforms in the ${\bf 70}$ representation of $SU(8)$ and obeys the self-duality condition
\be \bar{W}^{abcd} = {1 \over 4 !} \epsilon^{abcdefgh} W_{efgh} \ee 
and the differential constraint
\be \bar{D}_{\dot \a}^a W_{bcde} + {4 \over 5} \delta^a_{[b} \bar{D}^f_{\dot  \a} W_{cde]f} = 0 \ . \label{diffcon4D} \ee
The $\theta^0$ component of the superfield gives the $70$ scalar fields of the ${\cal N }=8$ supergravity multiplet. 
The differential constraint ensures that the components with one or more fermionic coordinates give the fields listed in Table 
\ref{Table_GM} without any extra degrees of freedom. Closure of the supersymmetry algebra requires that 
all fields in the  expansion of $W_{abcd}$ obey 
free-field equations of motion.

In the following sections we will introduce analogous constrained superfields 
for the fields of maximal supergravity in three and six dimensions. 
 
\renewcommand{\theequation}{3.\arabic{equation}}
\setcounter{equation}{0}

\bigskip 

\section{Unitary representations of $OSp(16|4,\mathbb{R})$ and ${\cal N}=16$  super\-gravity in three dimensions \label{3dconformalgroup}}

\subsection{Coherent states of the positive energy unitary representations of
$Sp(4,\mathbb{R})$ and conformal fields in three dimensions }

As illustrated in detail in Section~\ref{4dconformalgroup}, conformal
fields in $d$ dimensions correspond to coherent states labeled by the coordinate $d$-vector. They
are of the form
\[
e^{-i x_{\mu} P^{\mu}} T |\Omega\rangle
\]
where $T$ is the intertwiner to the non-compact basis and $|\Omega\rangle $ is the lowest energy
irreducible representation of the maximal compact subgroup $SO(d)\times SO(2)$ of the conformal group $SO(d,2)$.
The lowest energy irreducible representation $|\Omega\rangle$ is obtained by the action of creation operators on
the  Fock vacuum $|0\rangle$  in the compact (unitary) basis. The intertwiner $T$ can be
used to convert the oscillators in the compact basis to covariant oscillators in the non-compact basis acting on the "covariant Fock vacuum" $T|\Omega\rangle$.
The coherent state basis makes the analysis of dimensional reduction of a conformal field or
of a supermultiplet of fields very simple both in the compact unitary basis or in the non-compact
covariant basis.

For example in the non-compact picture consider the above coherent state at the origin $x_\mu=0$,
which is simply $T|\Omega\rangle$ which transform irreducibly under  the Lorentz group
$SO(d-1,1)$ with a definite scale  dimension $\ell$. One decomposes this irreducible representation with respect to
the Lorentz group of the lower dimension and restrict the coordinate vector $x_\mu$  to lie in
the lower dimensional Minkowski subspace. Since the entire coordinate dependence comes from
the action of translation generator  $e^{-i x_\mu P^\mu}$ on $T|\Omega \rangle$ this
decomposition yields directly the fields in the lower dimension. The only subtlety comes in
identifying the states that correspond to derivatives (descendants) of (primary) conformal fields
in the lower dimension. This identification is equivalent to dualization of the fields in the lower
dimension modulo the caveat that the oscillator method always yields states corresponding
to gauge invariant field strengths as opposed to gauge potentials.

\subsection{$SO(3,2) \approx Sp(4,\mathbb{R})$ representations via the oscillator method}

The conformal group of three-dimensional Minkowskian spacetime is  $SO(3,2)$ whose covering group is $Sp(4,\mathbb{R})$.
The compact  three-grading  of the
Lie algebra of ${Sp}(4,\mathbb{R}) \approx
{SO}(3,2)$ is determined by  its maximal compact
subalgebra ${SU}(2) \oplus {U}(1)$,
\[ {SO}(3,2) = A_{ij} \oplus U^i_j \oplus A^{ij} \ , \]
where $U^i_j$ denotes the $SU(2)\times U(1)$ generators. The non-compact three-grading is determined by the dilatation generator $\mathcal{D}$
\[ SO(3,2) = P_\mu  \oplus ( M^\mu_\nu + \mathcal{D})  \oplus K_\mu \]
where $\mu , \nu =0,1,2$ and $P_\mu$ ,  $M^\mu_\nu $ and $ K_\mu$ denote translation, Lorentz and special conformal generators, respectively. 

To construct the positive-energy UIRs of
$Sp(4,\mathbb{R})$, one introduces an arbitrary number of twistorial bosonic
oscillators.  However, unlike  the 
case of $SU(2,2)$ or of $SO^*(8)$, where one has to choose an
even number of oscillators $a_i(1),\dots,a_i(P)$;
$b_i(1),\dots,b_i(P)$ to realize the Lie algebra, here one also has the freedom of choosing an
odd number. In particular one can realize the Lie algebra of $Sp(4,\mathbb{R})$ in terms of a single set of twistorial bosonic oscillators, which lead to the famous singleton representations of Dirac \cite{Dirac}.

Let  $a_i(R)$, $b_i(R)$ be a set of bosonic annihilation operators with their
hermitian conjugate creation operators $a^i(R)$, $b^i(R)$, which
transform covariantly and contravariantly, respectively, under $SU(2)$. Let $i=1,2$ and $R=1,\dots,P$ where $P$ is 
the number of colors or generations of  oscillators. 
In addition,  let  $c_i$ and
its conjugate $c^i$ be a single set of such oscillators. They
satisfy the commutation relations,
\begin{equation}
\left[ a_i(R) , a^j(S) \right] = \delta_i^j \delta_{RS} \,, \qquad
\left[ b_i(R) , b^j(S) \right] = \delta_i^j \delta_{RS} \,, \qquad
\left[ c_i , c^j \right] = \delta_i^j \quad \,,
\label{U2commutators4x7}
\end{equation}
while all the other commutators vanish. The vacuum state $\ket{0}$ is
annihilated by all $a_i(R)$ and $b_i(R)$
as well as by all $c_i$.

The singleton realization of $Sp(4,\mathbb{R})$ is given by the following bilinears of  $c_i$ and $c^i$,
\begin{equation}
{M^i}_j =  \frac{1}{2}
           \left( c^i c_j + c_j c^i \right) 
\quad , \quad           
A_{ij} =  c_i c_j 
\quad , \quad
A^{ij} =  c^i c^j \, .
\label{Sp4Rsingleton}
\end{equation}
For the above realization of $Sp(4,\mathbb{R})$, there exist only two lowest energy irreducible representations of $SU(2)$ that are annihilated by $A_{ij}$. They are the Fock vacuum $| 0 \rangle$  and the one particle excitation $a^i |0\rangle$. These lowest energy irreducible representations determine the scalar and spinor singleton representations of $Sp(4,\mathbb{R})$. The corresponding coherent states
\eq
e^{i P_\mu x^\mu} T |0\rangle \equiv | \phi(x) \rangle
\en
and 
\eq
e^{i P_\mu x^\mu } T c^i|0\rangle \equiv | \psi^\alpha(x)\rangle
\en 
describe massless scalar and spinor conformal fields in three dimensions. The intertwiner operator $T$ converts oscillators transforming covariantly under the $SU(2)$ subgroup of $Sp(4,\mathbb{R})$ into oscillators transforming 
covariantly with respect to the Lorentz group $SU(1,1)$,
\eqn
\lambda^\alpha = T c^i T^{-1} \ ,
\qquad
\mu_\alpha = T c_i T^{-1} \ ,
\enn
where $\alpha, \beta =1,2$. The masslessness follows from the fact that the Poincar\'e mass 
operator $P_\mu P^\mu$ vanishes identically for the singleton irreducible representations. When restricted 
to the Poincar\'e subgroup these representations coincide 
with  the two massless representations  of Poincar\'e group in three dimensions that were classified in \cite{Binegar:1981gv}. They are the only massless representations of the Poincar\'e group in three dimensions.  

General positive energy unitary representations  of the three-dimensional conformal group are obtained by realizing its Lie algebra    as
bilinears of an arbitrary set of twistorial  bosonic oscillators in the following manner,
\begin{equation}
\begin{split}
{U^i}_j &= \mathbf{a}^i \cdot \mathbf{a}_j +
           \mathbf{b}_j \cdot \mathbf{b}^i +
           \frac{\epsilon}{2}
           \left( c^i c_j + c_j c^i \right) \ , \\
A_{ij} &= \mathbf{a}_i \cdot \mathbf{b}_j +
          \mathbf{a}_j \cdot \mathbf{b}_i +
          \epsilon ~ c_i c_j \ , \\
A^{ij} &= \mathbf{a}^i \cdot \mathbf{b}^j +
          \mathbf{a}^j \cdot \mathbf{b}^i +
          \epsilon ~ c^i c^j \,,
\label{Sp4Rgenerators}
\end{split}
\end{equation}
where $\epsilon = 0$ ($\epsilon = 1$) if the number of oscillators $n$
is even (odd) and the dot
product denotes summation over the generation or color index. The generators satisfy \cite{mgnw,Gunaydin:1988kz}
\begin{equation}
\left[ A_{ij} , A^{kl} \right] = \delta_i^k {U^l}_j + \delta_i^l {U^k}_j
     + \delta_j^k {U^l}_i + \delta_j^l {U^k}_i \,.
\end{equation}
${U^i}_j$ form the maximal compact subalgebra
${U}(2)$ of ${Sp}(4,\mathbb{R})$. The ${U}(1)$ charge which
defines the compact three-grading  is given by the trace ${U^i}_i$, 
\begin{equation}
E = \frac{1}{2} {U^i}_i = \frac{1}{2} N_B + P +  \frac{\epsilon}{2} 
\label{E47}
\end{equation}
and corresponds to the conformal Hamiltonian.  
The positive energy  UIRs of
$Sp(4,\mathbb{R}) $ can be constructed by choosing a
set of states $\ket{\Omega}$ that transforms irreducibly under $U(2)$
and is annihilated by all the grade $-1$ generators. Repeated action on these $\ket{\Omega}$ by the grade $+1$
generators gives the "particle basis" of the positive energy
UIRs. Poincar\'e mass does not vanish for the positive energy unitary representations of $Sp(4,\mathbb{R})$ 
 when the number $(2P + \epsilon) >1$. Hence, the conformal fields defined by the corresponding coherent states are not massless; this is consistent with the fact that there exist only two massless representations of the Poincar\'e group in $d=3$.


\subsection{$SO(16)$ representations via the oscillator method}

$SO(16)$ has a three-grading structure with respect to its subgroup
$U(8)$. To realize its Lie algebra we consider  fermionic annihilation and creation
operators that transform as $\overline{\mathbf{8}}$ and $\mathbf{8}$
representations of $U(8)$ satisfying the canonical anti-commutation relations,
\begin{equation}
\left\{ \alpha_\kappa(R) , \alpha^\rho(S) \right\} =
   \delta_\kappa^\rho \delta_{RS} \,,
\qquad
\left\{ \beta_\kappa(R) , \beta^\rho(S) \right\} =
   \delta_\kappa^\rho \delta_{RS} \,,
\qquad
\left\{ \gamma_\kappa , \gamma^\rho \right\} =
   \delta_\kappa^\rho \quad \mbox{(if present)}
\,, \label{U8commutators4x7}
\end{equation}
where 
$R=1,\dots,P$ and $\kappa,\rho=1, \dots, 8$.
 Once again, the vacuum
state $\ket{0}$ is annihilated by all the annihilation operators
$\alpha_\kappa(R)$, $\beta_\kappa(R)$ and $\gamma_\kappa$ (if present)
for all values of $\kappa$ and $R$.
The singletonic realization of $SO(16)$ is achieved in terms of a single set of oscillators,
\begin{equation}
{M^\kappa}_\rho = \frac{1}{2} \left( \gamma^\kappa \gamma_\rho -
        \gamma_\rho \gamma^\kappa \right) 
\quad,\quad
A_{\kappa\rho} =  \gamma_\kappa \gamma_\rho 
\quad,\quad
A^{\kappa\rho} =  \gamma^\kappa \gamma^\rho \, ,
\label{SO16singleton}
\end{equation}
where $M^\kappa_{~\rho}$ are the generators of the $U(8)$ subgroup. This realization 
leads to two irreducible representations of $SO(16)$: the spinor ${\bf 128}_s$ formed 
by the vacuum vector and the even excitations and the spinor ${\bf 128}_c$ formed by 
odd excitations. 

The general realization of Lie algebra of ${SO}(16)$ as  bilinears of
the above  fermionic oscillators is as follows,
\begin{equation}
\begin{split}
{M^\kappa}_\rho &= \boldsymbol{\alpha}^\kappa \cdot \boldsymbol{\alpha}_\rho -
     \boldsymbol{\beta}_\rho \cdot \boldsymbol{\beta}^\kappa +
      \frac{\epsilon}{2} \left( \gamma^\kappa \gamma_\rho -
        \gamma_\rho \gamma^\kappa \right) \ , \\
A_{\kappa\rho} &= \boldsymbol{\alpha}_\kappa \cdot \boldsymbol{\beta}_\rho -
     \boldsymbol{\alpha}_\rho \cdot \boldsymbol{\beta}_\kappa +
     \epsilon ~ \gamma_\kappa \gamma_\rho \ ,\\
A^{\kappa\rho} &= \boldsymbol{\alpha}^\kappa \cdot \boldsymbol{\beta}^\rho -
     \boldsymbol{\alpha}^\rho \cdot \boldsymbol{\beta}^\kappa +
     \epsilon ~ \gamma^\kappa \gamma^\rho \ .
\label{SO16generators}
\end{split}
\end{equation}
where $\epsilon =0$ or $\epsilon=1$ depending on whether we have an even or odd number of oscillators. 
The generators satisfy
\begin{equation}
\left[ A_{\kappa\rho} , A^{\lambda \sigma} \right] =
    \delta_\kappa^\lambda {M^\sigma}_\rho -
    \delta_\kappa^\sigma {M^\lambda}_\rho -
    \delta_\rho^\lambda {M^\sigma}_\kappa +
    \delta_\rho^\sigma {M^\lambda}_\kappa \,.
\end{equation}
 The
${U}(1)$ charge that determines the three-grading is given by 
\begin{equation}
C = \frac{1}{2} {M^\kappa}_\kappa = \frac{1}{2} N_F - 2P - \epsilon \,.
\label{C47}
\end{equation}

The set of vectors $\ket{\Omega}$ that transform
irreducibly under $U(8)$ and are annihilated by $A_{\kappa\rho}$ lead to irreducible representations 
of $SO(16)$ by the action of the grade $+1$ generators $A^{\kappa\rho}$.


\subsection{Unitary representations of $OSp(16|4,\mathbb{R})$ via the oscillator method
\label{Osposcillators3d}}

The superalgebra ${OSp}(16|4,\mathbb{R})$ has a three-grading with
respect to its compact sub-superalgebra ${U}(2|8)$. 
To construct the positive energy UIRs of $OSp(16|4,\mathbb{R})$,  consider  the
$U(2|8)$-covariant super-oscillators defined as follows,
\begin{equation}
\begin{split}
\xi_A(R) &= \left( \begin{matrix} a_i(R) \cr \alpha_\kappa(R)
            \cr \end{matrix} \right) \,, \qquad
\xi^A(R) = \xi_A(R)^\dag = \left( \begin{matrix} a^i(R)
           \cr \alpha^\kappa(R) \cr
           \end{matrix} \right) \ , \\
\eta_A(R) &= \left( \begin{matrix} b_i(R) \cr \beta_\kappa(R)
             \cr \end{matrix} \right) \,, \qquad
\eta^A(R) = \eta_A(R)^\dag = \left( \begin{matrix} b^i(R)
            \cr \beta^\kappa(R) \ , \cr
           \end{matrix} \right) \\
\zeta_A &= \left( \begin{matrix} c_i \cr \gamma_\kappa
           \cr \end{matrix} \right) \,, \qquad
\zeta^A = {\zeta_A}^\dag = \left( \begin{matrix} c^i
          \cr \gamma^\kappa \cr
          \end{matrix} \right) \ . \\
\end{split}
\end{equation}
where $i=1,2$ , $\kappa=1,\dots,8$ and $R=1,\dots,P$. The
oscillators $a$, $b$, $\alpha$ and $\beta$ as well as  $c$ and  $\gamma$ satisfy the usual (anti)commutation relations discussed in the previous subsections; it is easy to see that they imply that
\begin{equation}
\left[ \xi_A(R) , \xi^B(S) \right\} = \delta_A^B \delta_{RS} \,, \qquad
\left[ \eta_A(R) , \eta^B(S) \right\} = \delta_A^B \delta_{RS} \,, \qquad
\left[ \zeta_A , \zeta^B \right\} = \delta_A^B \,.
\end{equation}

The Lie superalgebra
${OSp}(16|4,\mathbb{R})$ can then be realized as follows:
\bea
{M^A}_B &=& {\boldsymbol{\xi}}^A \cdot {\boldsymbol{\xi}}_B +
           (-1)^{(\mathrm{deg} A)(\mathrm{deg} B)} \boldsymbol{\eta}_B \cdot
                 \boldsymbol{\eta}^A + {\epsilon \over 2} \left( \zeta^A \zeta_B +
           (-1)^{(\mathrm{deg} A)(\mathrm{deg} B)} \zeta_B \zeta^A \right) \ , \no \\
A_{AB} &=& \boldsymbol{\xi}_A \cdot \boldsymbol{\eta}_B + \boldsymbol{\eta}_A \cdot \boldsymbol{\xi}_B +
          \epsilon \ \zeta_A \zeta_B \ , \no \\
A^{AB} &=& \boldsymbol{\eta}^B \cdot \boldsymbol{\xi}^A + \boldsymbol{\xi}^B \cdot \boldsymbol{\eta}^A +
          \epsilon \ \zeta^B \zeta^A \ .
\label{OSp164Rgenerators2}
\eea
${M^A}_B$ generate the 
subalgebra  ${U}(2|8)$ and $A_{AB}$ and $A^{AB}$ extend it to
the full ${OSp}(16|4,\mathbb{R})$ superalgebra. The Abelian
${U}(1)$ charge which defines the above three-grading is
\begin{equation}
E + C = \frac{1}{2} {M^A}_A = \frac{1}{2} \left( N_B + N_F \right) -
        P -  \frac{\epsilon}{2} \  .
\end{equation}

Given this super-oscillator realization, the construction of 
positive energy UIRs of $OSp(16|4,\mathbb{R})$ proceeds by 
first choosing a set of states $\ket{\Omega}$ in the Fock space that 
transforms irreducibly under $U(2|8)$ and is annihilated by 
the grade $-1$ generators and then acting on it with the grade $+1$ generators. 
This generates an infinite set of states that form the particle basis 
of a positive energy UIR of $OSp(16|4,\mathbb{R})$.

\subsection{Supersingletons of $OSp(16|4,\mathbb{R})$ and fields of maximal supergravity in three dimensions}
When one reduces maximal supergravity to three dimensions, all the dynamical bosonic 
and fermionic fields can be  dualized to scalar and spinor fields and the resulting theory can 
be written as an ${\cal N}=16$ supersymmetric sigma model \cite{Marcus:1983hb,Nicolai:2000sc} with 
the scalar manifold 
\[ 
\mathcal{M}_3 = \frac {E_{8(8)}}{SO(16)} \, . 
\]
The scalar and spinor fields transform irreducibly as ${\bf 128}_s$ and ${\bf 128}_c$ of $SO(16)$, respectively. 
As in four dimensions one can  fit the 128 massless scalar fields and the 128 massless spinor 
fields into a conformal supermultiplet with $R$-symmetry $SO(16)$. This supermultiplet is the supersingleton 
of $OSp(16|4,\mathbb{R})$ obtained by considering a single set of superoscillators ({\it i.e.} 
the realization (\ref{OSp164Rgenerators2}) for $P=0$ and $\epsilon=1$),
\eqn
{M^A}_B =  \frac{1}{2} \left( \zeta^A \zeta_B +
           (-1)^{(\mathrm{deg} A)(\mathrm{deg} B)} \zeta_B \zeta^A \right)
\quad , \quad          
A_{AB} = \zeta_A \zeta_B 
\quad , \quad 
A^{AB} =   \zeta^B \zeta^A \,.
\label{OSp164Rgenerators}
\enn
 In this case there exist two possible lowest energy irreducible representations of $U(2|8)$ annihilated by $A_{AB}$ , namely the Fock vacuum $|0\rangle$ and the one superparticle state $\zeta^A|0\rangle$. Fock vacuum leads to a supermultiplet which in the coherent state basis can be decomposed as 
 \eq 
 {\bf 128}_s |\phi(x)\rangle \oplus {\bf 128}_c |\psi^\alpha(x)\rangle  \, .
 \en
 The $U(2|8)$ irreducible representation $\zeta^A|0\rangle$ leads to the  supermultiplet 
  \eq 
  {\bf 128}_c |\phi(x)\rangle \oplus {\bf 128}_s |\psi^\alpha(x)\rangle \, .
  \en
  Thus the fields of maximal Poincar\'e supergravity can be identified with the supersingleton multiplet of ${\cal N}=16$ conformal superalgebra $OSp(16|4,\mathbb{R})$ defined by the vacuum vector as the lowest weight state. 
  
  Let us now show that the doubleton supermultiplet of $SU(2,2|8)$ corresponding to the fields of $d=4$ maximal supergravity reduces to the supersingleton multiplet of $OSp(16|4,\mathbb{R})$ under dimensional reduction.
  In going from four to three dimensions the four-dimensional conformal group $SU(2,2)$ goes over to its subgroup $Sp(4,\mathbb{R})$. The maximal compact subgroup $SU(2)_L\times SU(2)_R\times U(1)$ of $SU(2,2)$ gets then restricted to the diagonal subgroup $SU(2)\times U(1)$. Thus in going down to three dimensions we must identify the spinors of $SU(2)_L$ and $SU(2)_R$,
  \eqn
  a_i \cong b_i := c_i \ , \qquad
  a^i \cong b^i := c^i \ .
  \enn
Hence
\eq
a_i b_j \rightarrow c_ic_j \hphantom{\ .} \ , \qquad
a^ib^j \rightarrow c^i c^j \hphantom{\ .} \ , \qquad 
a^ia_j \cong b^ib_j \rightarrow c^ic_j  \ .
\en
 We get a singletonic realization of $Sp(4,\mathbb{R})$ in three dimensions. 
 Then the coherent states describing the massless conformal fields in four dimensions reduce to scalars and spin-${1 \over 2}$ 
fields in three dimensions as tabulated in Table \ref{Table_3d}.
 \begin{table}[ht]
\begin{center}
\begin{tabular}{|c|c|}
\hline
{  $d=4$ Fields} & { $d=3$ Fields}
\\ \hline
$\phi^{[ABCD]}(x) $ & $\phi^{[ABCD]}(x) $
\\ \hline
$\lambda_{\alpha}^{[ABC]}(x)$ & $\psi_{\alpha}^{[ABC]}(x)$
\\ \hline
$\lambda_{\dot{\alpha}[ABC]}(x)$ & $\psi_{\alpha [ABC]}(x)$
\\ \hline
$ h_{(\alpha\beta)}^{[AB]}(x)$ & $ \partial_{\a \b} \phi^{[AB]}(x)$
\\ \hline
$h_{(\dot{\alpha}\dot{\beta})[AB]}(x)$& $\partial_{\a \b} \phi_{[AB]}(x)$ 
\\ \hline
$ \psi_{(\alpha\beta\gamma)}^{A}(x)$ & $\partial_{(\a \b} \psi_{\gamma)}^{A}(x) $
\\ \hline
$\psi_{(\dot{\alpha}\dot{\beta}\dot{\gamma}) A}(x)$ &$ \partial_{(\a \b} \psi_{\gamma)A}(x)$
\\ \hline
$R_{(\alpha\beta\gamma\delta)}(x)$ & $\partial_{(\a \b} \partial_{\gamma \delta)} \phi^+(x)$
\\ \hline
$ R_{(\dot{\alpha}\dot{\beta}\dot{\gamma}\dot{\delta} )}(x)$ & $\partial_{(\a \b} \partial_{\gamma \delta)}  \phi^-(x)$ 
\\ \hline
\end{tabular}
\medskip
\caption{\small \label{Table_3d} 
Three-dimensional
 decomposition of the CPT-self-conjugate doubleton supermultiplet of four-dimensional superconformal 
group $SU(2,2|8)$  under dimensional reduction. $A,B,C,..$ are the $SU(8)$ indices. 
Four-dimensional field-strengths yield scalars with derivatives in three dimensions.
}
\end{center}
\end{table}\\
We see that we get 128 massless scalar fields 
transforming in the ${\bf 128}_s$ of $SO(16)$ and 128 massless spin ${1 \over 2}$ fields transforming in ${\bf 128}_c$ of $SO(16)$.  
They can be identified with the fields of maximal ${\cal N}=16$ supergravity in three dimensions. 

\subsection{Three-dimensional constrained on-shell superfields}
 
We will now show that the fields of three-dimensional ${\cal N}=16$ supergravity can be conveniently 
organized in an on-shell superfield. We can introduce a suitable three-dimensional superspace with coordinates
\cite{Samtleben:2009ts},
\be \big( x^{\a \b}, \theta^{\a \hat a} \big) \qquad \mathrm{with} \quad \a,\b=1,2 \quad \hat a=1,2, \dots, 16  \ . \ee 
The bosonic coordinates $x^{\a \b}$ are real symmetric matrices in the spinor indices. 
The fermionic coordinates $\theta^{\a \hat a}$ are Majorana Lorentz spinors and transform in the 
16-dimensional vector representation of $SO(16)$. All the $R$-symmetry indices will be hatted throughout this section.
One can introduce the superspace covariant derivatives,
\be 
D_{\a \hat a} = \p_{\a \hat a} + i \theta^\b_{\hat a} \partial^{\vphantom{A}}_{\a \b} \ . 
\ee
With this definition, the covariant derivatives obey the relation
\be \{ D_{\a \hat a}, D_{\b \hat b} \} = 2 i \delta_{\hat a \hat b} \partial^{\vphantom{A}}_{\a \b} \ .
\label{comm-D-3d}  \ee    
We can then introduce the superfield
\be 
\Phi_{\ha} \big( x^{\a \b}, \theta^{\a \hat a} \big)   
\ee 
which transforms as a Lorentz scalar and as a (real) ${\bf 128}_s$ spinor of $SO(16)$. If $\ha=1,2, \dots 256$ 
is an $SO(16)$ spinor index, the superfield needs to obey the $SO(16)$ chirality condition
\be 
\Gamma_{17} \Phi = + \Phi \ .
\ee 
In order to have the right number of degrees of freedom in the superfield, 
we also need to impose the differential constraint,
\be D_\a^{\hat a} \Phi_\ha = {1 \over 16} \big( \Gamma^{\hat a} \Gamma^{\hat b} \big)_{\ha}^{\ \hb} D^{\hat b}_\a \Phi_\hb  \ ,
\label{diffconstraint3d}  \ee
where $\Gamma^{\hat a}$ are  $SO(16)$, $256\times 256$ gamma matrices.
We can define an additional superfield
\be 
\Psi_{\a \ha} = {1 \over 16} \big( \Gamma^{\hat a} \big)^{\ \hb}_\ha D_{\a}^{\hat a} \Phi_\hb  \ .
\ee
With this definitions, all the $128$ bosonic fields belong to the $\theta^0$ term of the expansion of the superfield $\Phi_\ha$ 
and all the $128$ fermionic fields belong to the $\theta^0$ term of the superfield $\Psi_{\a \ha}$. 

Due to the presence of the gamma matrix in the definition above, if the superfield $\Phi_\ha$ transforms in the ${\bf 128}_s$
representation of $SO(16)$, the superfield $\Psi_{\a \ha}$ will transform in the ${\bf 128}_c$ representation and vice-versa.
We can then rewrite the differential constraint (\ref{diffconstraint3d}) as
\be 
D_\a^{\hat a} \Phi_\ha = \big(  \Gamma^{\hat a}  \big)^{\ \hb}_\ha \Psi_{\a \hb} \ .
\label{diffconstraint3d-Phi} 
\ee
Using the relations (\ref{comm-D-3d}) and (\ref{diffconstraint3d}), 
we obtain a differential constraint for the superfield $\Psi_{\a \ha}$,
\be 
D_\a^{\hat a} \Psi_{\b  \ha} = i \big(  \Gamma^{\hat a}  \big)^{\ \hb}_\ha \p_{\a \b} \Phi_{\hb} \ .
\label{diffconstraint3d-Psi} 
\ee
in agreement with the constraints imposed in \cite{Greitz:2011vh}. 
Note that analogous constraints are also imposed in the ${\cal N}=8$ case \cite{Ferrara:2000dv}.
The expansions of both superfields contain terms up to $\theta^{32}$. However, 
due to the relations (\ref{diffconstraint3d-Phi}) and (\ref{diffconstraint3d-Psi}), 
all terms with more than one fermionic coordinate do not contain any new degrees of freedom.
 
Moreover, taking the $\theta^0$ components of 
(\ref{diffconstraint3d-Phi}) and (\ref{diffconstraint3d-Psi}) we obtain the supersymmetry transformations of the component 
fields of the ${\cal N}=16$ multiplet,
\bea 
\delta_\eta \phi_\ha &=& 
\eta^{\hat a \ha} \big(  \Gamma^{\hat a}  \big)^{\ \hb}_\ha \psi_{\a \hb} \\ 
\delta_\eta \psi_{\b  \ha} &=& 
i \eta^{\hat a \ha} \big(  \Gamma^{\hat a}  \big)^{\ \hb}_\ha \p_{\a \b} 
\phi_{\hb} \, ,
\eea 
where $\phi_\ha$ and $\psi_{\a \ha}$ are the bosonic and fermionic fields of the supermultiplet. 
  
\renewcommand{\theequation}{4.\arabic{equation}}
\setcounter{equation}{0}

\bigskip

\section{Unitary representations of $OSp(8^*|8)$ and maximal super\-gravity in six dimensions \label{6dconformalgroup}}

\subsection{Coherent states of the positive energy unitary representations of
$SO^{*}(8)$ and conformal fields in six dimensions}

Our main goal in this section is to construct the six-dimensional counterpart of the CPT-self-conjugate
doubleton supermultiplet of $SU(2,2|8)$ and discuss how it may be related to maximal supergravity
in six dimensions. To this end we shall first review, following \cite{gnw,mgst,FGT}, 
the oscillator construction of the positive
energy  unitary representations of the conformal group in six dimensions and its supersymmetric
extensions.

The generators of the conformal group $SO(6,2)$ in $d = 6$
 satisfy the commutation relations
\eq
\left[ M_{ab} , M_{cd} \right] = i ( \eta_{bc} M_{ad} - \eta_{ac} M_{bd} - \eta_{bd} M_{ac}
+ \eta_{ad} M_{bc} ), \label{SO62}
\en
where $a,b,c,d = 0,1,\dots, 7$ and $\eta_{a b}=\textrm{diag}(+,-,-,-,-,-,-,+)$.
$M_{\mu \nu}$, with $\mu,\nu=0,1, \dots, 5$, are the generators of the Lorentz subgroup $SO(5,1)$. 
The dilatation generator $D$,  the generators of translations $P_{\mu}$ and
the special conformal generators $K_{\mu}$ are related to the generators above by
\eq
M_{\mu 6} = \frac{1}{2} (P_{\mu} - K_{\mu}), \quad M_{\mu 7} = \frac{1}{2} (P_{\mu}
 + K_{\mu}), \quad M_{67} = -D,
\en
The covering group of the conformal group $SO(6,2)$ is $Spin(6,2)$ which is isomorphic to $SO^*(8)$ and the
  covering group of the Lorentz group $SO(5,1)$ is $SU^*(4)$. The rotation subgroup $SO(5)$
   (or its covering group $USp(4)$) is generated by $M_{\mu \nu}$, with $\mu,\nu = 1,2,\ldots,5$.

The generators
$M_{mn}$ ($m,n = 1,2,\ldots,6$) generate the compact subgroup $SU(4) $ and the $U(1)_E$ generator $E \equiv M_{07}$ is the  conformal Hamiltonian.

The Lie algebra of the conformal group $SO(6,2)$ has a three-graded decomposition with
respect to its maximal compact subalgebra ${\cal L}^{0} = SU(4) \times U(1)_{E}$,
\eq
{SO}(6,2) = \mathcal{L}^{-} \oplus \mathcal{L}^{0} \oplus \mathcal{L}^{+},
\en
where the  three-grading is determined by the conformal Hamiltonian  $ E = \frac{1}{2} ( P_{0} + K_{0} )$,
To construct positive
energy unitary representations of $SO^{*}(8)$, one realizes  the generators  as bilinears
 of an arbitrary number $P$ of pairs of bosonic annihilation
 (${\bf a}_{i}, {\bf b}_{j}$) and creation (${\bf a}^{i}, {\bf b}^{j}$) operators
(with $i,j = 1,2,3,4$), transforming in the fundamental representation of $SU(4)$ and its
 conjugate, respectively \cite{gnw,mgst,mgcs,ibmg,mgrs},
\eqn
A_{ij} & := & {\bf a}_{i} \cdot {\bf b}_{j} - {\bf a}_{j} \cdot {\bf b}_{i} \ , \nn \\
A^{ij} & := & {\bf a}^{i} \cdot {\bf b}^{j} - {\bf a}^{j} \cdot {\bf b}^{i} \ , \nn \\
M^{i}_{~j} & := & {\bf a}^{i} \cdot {\bf a}_{j} + {\bf b}_{j} \cdot {\bf b}^{i} \ . \label{CB}
\enn
The dot product denotes contraction with respect to the color or generation index, {\it i.e.} ${\bf a}_{i} \cdot {\bf b}_{j} := \sum_{R=1}^{P} a_i(R) b_j(R)$. 
The bosonic
 annihilation and creation operators $a^{i}(R) = a_{i}(R)^{\dag}$ and $b^{j}(R) =
 b_{j}(R)^{\dag}$ satisfy the usual canonical commutation relations
\eqn
\left[ a_{i}(R) , a^{j}(S) \right]  =  \delta_{i}^{~j} \delta_{RS}
\quad, \quad
\left[ b_{i}(R) , b^{j}(S) \right]  =  \delta_{i}^{~j} \delta_{RS},
\label{cancomrel6d}
\enn
where $i,j = 1,2,3,4$ and $R,S = 1,2,\ldots,P$.

$M^{i}_{~j}$ generate the maximal compact subgroup $U(4)$.
The conformal Hamiltonian is given by the trace  $M^i{}_i$
\eqn
Q_{B} & := & \frac{1}{2} M^{i}_{~i}  = \frac{1}{2} \left( N_{B} + 4 P \right),
\enn
where $N_{B} \equiv {\bf a}^{i} \cdot {\bf a}_{i} + {\bf b}^{i} \cdot {\bf b}_{i}$,
which is the bosonic number operator. We shall denote the eigenvalues of $Q_{B}$ as $E$.
The hermitian linear combinations of $A_{ij}$ and $A^{ij}$ are the non-compact generators of
 $SO(6,2)$ \cite{gnw,mgst,mgcs}.
Each lowest weight  (positive energy) UIR is
uniquely determined by a set of states transforming in the lowest energy irreducible representation  $| \Omega \rangle$ of  $SU(4) \times U(1)_{E}$ that
are annihilated by all the elements of $\mathcal{L}^{-}$ \cite{gnw,mgst}.\footnote{ Equivalently,  the lowest weight vector of the lowest energy irreducible representation of $SU(4)$ determines the UIR. Hence, by an abuse of terminology, we shall use interchangeably the terms "lowest weight vector" and "lowest energy irreducible representation". }
The
possible lowest weight vectors for $P = 1$, which are called doubleton representations \cite{mgst},  in
this compact basis,  are of the form
\eqn
| 0 \rangle, & & \nn \\
a^{i_1} | 0 \rangle & = & | \onebox \rangle, \nn \\
a^{( i_1} a^{i_2 )} | 0 \rangle & = & | \twobox \rangle, \nn \\
~ & \vdots & ~ \nn \\
a^{( i_1} a^{i_2} \ldots a^{i_n )} | 0 \rangle & = & | \marcnbox \rangle, 
\enn 
together with
those obtained by interchanging $a$-type oscillators with $b$-type oscillators and together with the
state \eq a^{(i} b^{j)} | 0 \rangle = | \twobox \rangle. \en
These lowest energy irreducible representations
$| \Omega \rangle$ - or  doubleton UIRs - of $SO^{*}(8)$ all transform in the symmetric
tensor representations of $SU(4)$.

Conformal fields in six dimensions transform covariantly under the Lorentz group
$SU^*(4)=Spin(5,1)$ and have definite conformal dimension.
Similarly to the four-dimensional conformal group discussed in
Section~\ref{4dconformalgroup},
the conformal group $SO(6,2)$ also has a non-compact three-graded structure 
with respect to its subgroup $SU^{*}(4) \times \mathcal{D}$,
where $\mathcal{D}$ is the dilation generator~\cite{mgst,mgm}.
Under the action of  $SO(6,2)$   on the (conformal compactification of) six-dimensional
Minkowski spacetime, the stability group $\mathcal{H}$ of the origin $x^{\mu} = 0$
is the semi-direct product
\[
( SU^{*}(4) \times \mathcal{D} ) \circledS \mathcal{K}_6~~,
\]
where $\mathcal{K}_6$ represents the Abelian subgroup generated by the special conformal
generators $K_{\mu}$. Conformal fields in $d = 6$ live on the coset space $G/\mathcal{H}$ and
 are labeled by their transformation properties under the Lorentz group
$SU^*(4)$, their conformal (scale) dimension $l$ and certain matrices $\kappa_{\mu}$ that
describe their behavior under special conformal transformations $K_{\mu}$ \cite{mgst}, in complete parallel
the four-dimensional situation reviewed in Section~\ref{4dconformalgroup}
\cite{macksalam,GMZ2}.

Representations of the six-dimensional conformal algebra may be conveniently constructed in
terms of the twistorial spinor
\eq
\Psi(R) :=
\left( \begin{array}{c} a_{i}(R) \\ b^{j}(R) \end{array} \right), 
\en 
and its Dirac conjugate
 \eqn
\bar{\Psi}(R) & \equiv & {\Psi}^{\dagger}(R) \Gamma_{0}
=  \left( \begin{array}{cc} a^{i}(R) & - b_{j}(R) \end{array}
\right). \enn

The generators of $SO(6,2)$ can be written as bilinears of these   twistorial  oscillators
 \eq \bar{ \bf \Psi} \Sigma_{ab} { \bf \Psi} :=
\sum_{R=1}^{P} \bar{\Psi}(R) \Sigma_{ab} \Psi(R),
\en
and  satisfy the commutation relations
\eq \left[ \bar{\bf \Psi} \Sigma_{ab} {\bf \Psi} , \bar{\bf \Psi} \Sigma_{cd} { \bf \Psi}
\right] = \bar{\bf \Psi} \left[ \Sigma_{ab} , \Sigma_{cd} \right] {\bf \Psi},
\en
as a consequence of the the canonical commutation relations (\ref{cancomrel6d})
of the oscillators $a_i$ and $b_j$. The matrices $\Sigma_{ab}$ are defined in terms of the
six-dimensional Dirac matrices
$\Gamma_\mu$, with $\{\Gamma_\mu,\Gamma_\nu\}=2
\eta_{\mu\nu}$, as
\be
\Sigma_{\mu\nu} :=  \frac{i}{4} \left[ \Gamma_{\mu} , \Gamma_{\nu} \right],
\qquad
\Sigma_{\mu 6} :=  - \frac{1}{2} \Gamma_{\mu} \Gamma_{7},
\qquad
\Sigma_{\mu 7}  :=  \frac{1}{2} \Gamma_{\mu},
\qquad
\Sigma_{67} :=  \frac{i}{2} \Gamma_{7}~,
\ee
where $\Gamma_{7} = - \Gamma_{0} \Gamma_{1} \Gamma_{2} \Gamma_{3}
\Gamma_{4} \Gamma_{5}$. The $\Sigma$ matrices generate the eight-dimensional
"left-handed" spinor representation of the conformal  algebra $SO(6,2)$\footnote{Our
conventions for $\Gamma$ matrices follow \cite{FGT} and  are outlined in Appendix B.}.
The Lorentz covariant generators of the conformal algebra are
\eqn
M_{\mu \nu} & = &
\frac{i}{4} \bar{\bf \Psi} \left[ \Gamma_{\mu} ,
 \Gamma_{\nu} \right] { \bf \Psi} ,
 \qquad\quad~~
D =
- \frac{i}{2} \bar{{ \bf \Psi}} \Gamma_{7}  { \bf \Psi} , \nn \\
P_{\mu} & = & \frac{1}{2} \bar{{ \bf \Psi}}  \Gamma_{\mu} ( I - \Gamma_{7} )
 { \bf \Psi} ,
  \qquad
K_{\mu} =
\frac{1}{2} \bar{{ \bf \Psi}}  \Gamma_{\mu} ( I + \Gamma_{7} ) { \bf \Psi} ~~.
 \enn

The positive energy UIRs of
$SO^{*}(8)$ can be identified with conformal fields in six dimensions,  transforming
covariantly under the six-dimensional Lorentz group and with definite conformal dimension and
trivial special conformal parameters $\kappa_\mu$.
To establish this correspondence
consider the  operator
\eq
T := e^{ \frac{\pi}{2} \bar{{ \bf \Psi}}  \Sigma_{06}{ \bf \Psi}}
\label{6dintertwiner}
\en
which satisfies the following important relations :
\eqn
M_{mn} T & = & T M_{mn} \hspace{2cm} \mbox{for } m,n = 1,2,\ldots,5, \nn \\
i M_{m0} T & = & T  M_{m6} , \nn \\
i D T & = & T E , \nn \\
K_{\mu} T & = &  T \mathcal{L}^{-}, \label{UL_d6}
\enn
where $\mathcal{L}^{-}$ stands for certain linear combinations of the  operators $A_{ij}$.
Therefore, similarly to the intertwiner of the four-dimensional conformal algebra given in equation
(\ref{4dintertwiner}), $T$ in equation (\ref{6dintertwiner}) intertwines between
the generators $( M_{\mu \nu}, D )$ of Lorentz group and dilatations and the
generators $(M_{mn},E )$ of the maximal compact  subgroup $SU(4) \times U(1)$.
Thus, the transformation properties of the states $ |\Omega \rangle$ that make up  
the minimal energy irreducible representation  of $SU(4)$ coincide with the transformation properties of 
$T|\Omega \rangle$ under the Lorentz group $SU^*(4)$.

Therefore  every unitary lowest weight representation (ULWR) of $SO^{*}(8)$ can be
identified with a unitary representation of $SO(6,2)$ induced by a finite dimensional
irreducible representation of $SU^{*}(4)$ with a definite conformal dimension $l$ and
trivially realized $K_{\mu}$.\footnote{This  follows from the fact that
${\cal L}^{-} | \Omega \rangle = 0$ implies that
$K_{\mu} T | \Omega \rangle = 0$ \cite{mgst}.}
The states constructed above do not carry any position dependence, {\it i.e.} they
are located at $x^\mu=0$. As discussed in Section \ref{4dconformalgroup} in a four-dimensional context, 
a state at any other space-time point is generated by the
action of the translation operator:
\eq
e^{i x^{\mu} P_{\mu}} T | \Omega \rangle = | \Phi_{( d_{1}, d_{2}, d_{3} )}(x)
\rangle~,
\en
where $(d_{1}, d_{2}, d_{3})$ are the Dynkin labels. Thus, every irreducible ULWR of
$SO(6,2)$ corresponds to a conformal field that transforms covariantly under $SU^{*}(4)$
with a definite conformal dimension $l = -E$.
These coherent states correspond to  six-dimensional conformal fields.

The doubleton
representations of $SO^{*}(8)$ are constructed using a single pair ($P = 1$) of bosonic
oscillators.
The Poincar\'{e} mass operator \eq M^{2} = P_{\mu} P^{\mu} \en in six-dimensional Minkowski
spacetime vanishes identically for the doubleton representations and hence the corresponding
conformal fields are massless. Below we shall restrict ourselves to  the massless (doubleton)
representations of the
conformal group \footnote{ Recently it was shown that massless doubleton representations
correspond to the minimal unitary representation of $SO^*(8)$ and its deformations are
labelled by $SU(2)$ spin, which is the six-dimensional analog of helicity in four dimensions \cite{FG2010b}.}.

The $SU^{*}(4)$ covariant spinorial oscillators are obtained from the $SU(4)$ covariant oscillators
by the action of intertwining operator $T$. We will use hatted Greek letters for the
$SU^{*}(4)$  indices -- $\hat\alpha, \hat\beta = 1,2,3,4.$ With this notation, the $SU^{*}(4)$ covariant
spinorial oscillators are:
 \eqn
\lambda^{\hat{\alpha}1} & =\frac{1}{\sqrt{2}} T \left( \begin{array}{c} a^{1} \\ a^{2} \\ a^{3} \\ a^{4} \end{array} \right) T^{-1} 
= \frac{1}{\sqrt{2}}  \left( \begin{array}{c} a^{1} - b_3 \\ a^{2} + b_4 \\ a^{3} + b_1 \\ a^{4} - b_2 \end{array} \right) \ , 
\enn
\eqn
 \lambda^{\hat{\alpha}2}  & = \frac{1}{\sqrt{2}}T \left( \begin{array}{c} b^{1} \\ b^{2} \\ b^{3} \\ b^{4} \end{array} \right) T^{-1} 
= \frac{1}{\sqrt{2}}  \left( \begin{array}{c} b^{1} + a_3 \\ b^{2} - a_4 \\ b^{3} - a_1 \\ b^{4} + a_2 \end{array} \right) \ , \\
\eta_{\hat{\alpha} 1} & = \frac{1}{\sqrt{2}}T \left( \begin{array}{c} a_{1} \\ a_{2} \\ a_{3} \\ a_{4} \end{array} \right) T^{-1}
= \frac{1}{\sqrt{2}}  \left( \begin{array}{c} a_{1} + b^3 \\ a_{2} - b^4 \\ a_{3} - b^1 \\ a_{4} + b^2 \end{array} \right) \ , \\
 \eta_{\hat{\alpha} 2} &  = \frac{1}{\sqrt{2}}T \left( \begin{array}{c} b_{1} \\ b_{2} \\ b_{3} \\ b_{4} \end{array} \right) T^{-1}
= \frac{1}{\sqrt{2}}  \left( \begin{array}{c} b_{1} - a^3 \\ b_{2} + a^4 \\ b_{3} + a^1 \\ b_{4} - a^2 \end{array} \right) \ . 
  \label{stuvspinors}
\enn
As a consequence of equation~(\ref{cancomrel6d}), they satisfy canonical commutation relations
\eq
[ \eta_{\hat{\alpha} i} , \lambda^{\hat{\beta} j} ] = \frac{1}{2}\delta_{\hat{\alpha}}^{\hat{\beta}} \delta_i^j~~,
\en
where $i,j = 1, 2 $ and, as mentioned above, $ \hat{\alpha}, \hat{\beta}  =1,2,3,4 $.
Then we find that
\eqn
\left(\Sigma^\mu \right)^{\ha\hb} P_{\mu}  = 
- 2 \lambda^{\hat{\alpha} i} \lambda^{\hat{\beta}j} \epsilon_{ij}
\quad,\qquad
\left( \Sigma^{\mu} \right)_{\ha\hb} K_{\mu}  = 
- 2 \eta_{\hat{\alpha} i } \eta_{\hat{\beta} j } \epsilon^{ij}~~;
\enn
where  $\Sigma$-matrices in $d = 6$ are the analogs of Pauli
matrices $\sigma_{\mu}$ in $d = 4$.\footnote{The explicit form of the $\Sigma_{\mu}$ matrices
 is given in Appendix B.}
 \eq 
 \overline{\Sigma}_{\mu} = ( \Sigma_{0}, - \Sigma_{1}, - \Sigma_{2}, - \Sigma_{3},
 - \Sigma_{4}, - \Sigma_{5})~~.
 \en
The first equation above is similar to the one used in the six-dimensional spinor helicity formalism \cite{COC, DHS}.
The doubleton irreducible representations of $SO(6,2)$ and the corresponding conformal fields are listed in  Table \ref{6d_doubletons}.
Thus the massless conformal fields in six dimensions can be labelled as  symmetric tensors in the
spinor indices as $\Phi_{ABC...E}(x)$ corresponding to a Dynkin label $(n,0,0)$ of $SU^*(4)$. 
This shows that the massless  graviton, which transforms in the 20-dimensional representation 
of $SU^*(4)$ with Dynkin label $(0,2,0)$, cannot be a massless conformal field. Similarly, a 
massless vector field in six dimensions cannot be conformal since it transforms in the $(0,1,0)$ 
representation of $SU^*(4)$.

\begin{table}[ht]
\begin{center}
\begin{tabular}{|c|c|c|}
\hline
 { \small States of lowest energy }  & { \small $SU^*(4)$ Dynkin labels}  & { \small conformal dimension } \\
{ \small irreps of $SU(4)$} & { \small coherent states $| \Phi_{( d_{1}, d_{2}, d_{3} )}(x) \rangle$ } & $l$ \\ \hline \hline
$ | 0 \rangle$ & $| \Phi_{(0,0,0)}(x) \rangle$ & $-2$ \\ \hline
$a^{\hat{i}_{1}}  | 0 \rangle$ & $| \Phi_{(1,0,0)}(x) \rangle$ & $- \frac{5}{2}$ \\ \hline
$a^{( \hat{i}_{1}} a^{\hat{i}_{2} )}  | 0 \rangle$ & $| \Phi_{(2,0,0)}(x) \rangle$ & $-3$ \\
\hline
\vdots & \vdots & \vdots \\ \hline
$a^{( \hat{i}_{1}} \ldots a^{\hat{i}_{n} )}  | 0 \rangle$ & $| \Phi_{(n,0,0)}(x)
 \rangle$ & $- \frac{1}{2}(n+4)$ \\ \hline
$a^{( \hat{i}_{1}} b^{\hat{j}_{2} )}  | 0 \rangle$ & $| \Phi_{(2,0,0)}(x)
\rangle$ & $-3$ \\ \hline
\end{tabular}
\end{center}
\caption{ \small \label{6d_doubletons} 
States transforming in the lowest energy irreducible representation of $SU(4)$ in the compact basis with their $SU^*(4)$ Dynkin labels
and conformal (scale) dimensions. }
\end{table}

\subsection{Six-dimensional conformal superalgebras $OSp(8^*| 2N)$ \label{Osposcillators6d}}

Simple conformal superalgebras in six dimensions belong to the family $OSp(8^*| 2N)$
with the even subgroup $SO^*(8) \times USp(2N)$. They correspond to chiral $(N,0)$
supersymmetry since the odd generators belong to a spinorial representation of definite
chirality\footnote{ Actually due to triality properties of $SO(8)$, there exist three different
forms of these superalgebras which are in triality. }.

The superalgebra $OSp(8^*| 4)$ appearing at $N=2$ may be interpreted both as the
$\mathcal{N}=4$  extended $AdS$ superalgebra in $d=7$ or as the $(2,0)$ extended
conformal superalgebra with $32$ supercharges in six dimensions. The interacting
quantum theory of the $(2,0)$ doubleton supermultiplet of $OSp(8^*|4)$  is believed to be dual
to $M$-theory over $AdS_7 \times S^4$.

The supersymmetry generators
$Q_{\Gamma A }$ of $OSp(8^*| 2N)$  satisfy the  anticommutation relations  \cite{ckp}
\eqn
\left\{ Q_{\Gamma A}, Q_{\Delta B} \right\} = -\frac{1}{2} \left(
\Omega_{AB}
M_{\Gamma \Delta} + C_{\Gamma \Delta} U_{AB} \right)~~,
\enn
where $A,B,..=1,..,2N$ and $\Gamma, \Delta =1,...8$. $U_{AB}=U_{BA}$
are the $USp(2N)$ generators and $\Omega_{AB }=-\Omega_{BA}$ is the symplectic
invariant tensor. The tensor $C_{\Gamma\Delta}$ is the charge conjugation matrix
in $(6,2)$ dimensions and is symmetric \cite{ckp}.

The generators  of $USp(2N)$ satisfy
\eqn
\left[ U_{AB}, U_{CD} \right] = \Omega_{A(C}U_{D)B} + \Omega_{B(C}U_{D)A}~~.
\enn
The  commutation  relations of $SO(6,2)$ and $USp(2N)$ with the supersymmetry
generators,
\eqn
\left[ M_{ab}, Q_{\Gamma A} \right] =  \left( \Sigma_{ab}
\right)_{\Gamma}{}^{\Delta} Q_{\Delta A}, \nn \\
\left[ U_{AB}, Q_{C \Gamma} \right] = -\Omega_{C(A} Q_{B)\Delta}~~,
\enn
identify the supercharges $Q_{\Delta A}$ as the bi-fundamental representation of the bosonic
subgroup.

The superalgebra  $OSp(8^*| 2N)$ has a three-graded decomposition with
respect to its compact subsuperalgebra ${\cal L}^0=U(4|N)$,
\eq
OSp(8^*| 2N)= \mathcal{L}^{+} \oplus \mathcal{L}^{0} \oplus \mathcal{L}^{-},
\en
This three-grading restricts to the three-grading of $SO^*(8)$ with respect to $U(4)$ subgroup and to the three-grading 
of $USp(2N)$ with respect to its $U(N)$ subgroup.

The Lie superalgebra $OSp(8^*| 2N)$   is then  realized
as bilinears  of super-oscillators transforming in the fundamental representation of $U(4|N)$ and its conjugate, 
respectively. The superoscillators are defined as
\begin{equation}
\begin{split}
\xi_{\hat A}(R) &= \left( \begin{matrix} a_i(R) \cr \alpha_{x}(R)
            \cr \end{matrix} \right) \,, \qquad
\xi^{\hat A}(R) = \left( \begin{matrix} a^i(R)
           \cr \alpha^{ x}(R) \cr
           \end{matrix} \right) \\
\eta_{\hat A}(R) &= \left( \begin{matrix} b_i(R) \cr \beta_{x}
             \cr \end{matrix} \right) \,, \qquad
\eta^{ \hat A}(R) = \left( \begin{matrix} b^i(R)
            \cr \beta^{ x}(R) \cr
           \end{matrix} \right) \end{split}
\end{equation}
where $i=1,\dots, 4$ , $x=1,\dots,N$ and $R=1,\dots,P$. The
oscillators satisfy the usual graded commutation relations,
\begin{equation}
\left[ \xi_{\hat A}(R) , \xi^{\hat B}(S) \right\} = \delta_{\hat A}^{\hat B} \delta_{RS} \,, \qquad
\left[ \eta_{\hat A}(R) , \eta^{\hat B}(S) \right\} = \delta_{\hat A}^{\hat B} \delta_{RS} \,.
\end{equation}
Then, the grade $+1$, grade $0$ and grade $-1$ generators respectively have the following expressions
\bea
A^{\hat A \hat B} &:=& \boldsymbol{\eta}^{\hat B} \cdot \boldsymbol{\xi}^{\hat A} + \boldsymbol{\xi}^{\hat B} \cdot \boldsymbol{\eta}^{\hat A}, \no \\
{M^{\hat A}}_{\hat B} &:=& \boldsymbol{\xi}^{\hat A} \cdot \boldsymbol{\xi}_{\hat B} +
           (-1)^{(\mathrm{deg} {\hat A})(\mathrm{deg} \hat B )} \boldsymbol{\eta}_{\hat B} \cdot
                 \boldsymbol{\eta}^{\hat A}, \no \\
A_{\hat A \hat B} &:=& \boldsymbol{\xi}_{\hat A} \cdot \boldsymbol{\eta}_{\hat B} + \boldsymbol{\eta}_{\hat A} \cdot \boldsymbol{\xi}_{\hat B}. 
\label{OSp8NRgenerators}  
\eea 
The resulting unitary supermultiplets of $OSp(8^*| 2N)$  in the super-Fock space  
decompose into a finite set of positive energy irreducible representations of $SO^*(8)$ transforming in irreducible representations of $USp(2N)$.  For a single pair of super-oscillators  $P=1$ (doubletons)  the resulting representations correspond to  superconformal multiplets of massless fields in six dimensions.

\medskip

\begin{table}[hb]
\begin{center}
\begin{tabular}{|c|c|c|}
\hline
Conformal Field &  ${SU^{*}(4)}_{D}$ & $USp(4)$
\\ \hline

$\phi^{[AB]|}(x)$  & (0,0,0)       & 5     \\ \hline
$\lambda_{\hat{\alpha}}^{A} (x) $ & (1,0,0)
& 4 \\ \hline

$h_{(\hat{\alpha} \hat{\beta})}(x)$ & (2,0,0)       & 1     \\ \hline
\end{tabular}
\end{center}
\caption{\label{6d_doubleton} \small Doubleton supermultiplet of $OSp(8^*|4)$ corresponding to the lowest weight vector
$|\Omega\rangle = |0\rangle$, with  $SU^*(4)$ Dynkin labels and  dimension of the $USp(4)$  representations.  
The conformal fields in the first column are labelled by  
$SU^*(4)$ spinor indices $\hat{\alpha},\hat{\beta},\dots$ and $USp(4)$ indices $A,B,\dots$. The square bracket $[\,\cdots ]|$ denote an antisymmetric, symplectic-traceless tensor.}
\end{table}
The doubleton supermultiplet of \OS, which is defined
by the lowest weight vector $|\Omega\rangle = |0\rangle$,
is the massless $(2,0)$ conformal supermultiplet and is the analog of the
${\cal{N}}=4$ super Yang-Mills multiplet in $d=6$ \cite{gnw}.  The content of
the (2,0) supermultiplet is given in Table \ref{6d_doubleton}.

In Table \ref{USp(8)_doubleton} we give the field content of the doubleton supermultiplet of $OSp(8^*|8)$ with 
$R$-symmetry group $USp(8)$, namely the $(4,0)$ conformal supermultiplet. 
The oscillator construction of the positive energy unitary supermultiplets of $OSp(2M^*|2N)$
with the even subgroup $SO^*(2M) \times USp(2N)$ was given in \cite{gunasca}. For both the $(2,0)$ 
and the $(4,0)$ supermultiplets, the oscillator construction gives directly the gauge-invariant field strengths. 
The corresponding gauge potentials are discussed in Section \ref{secpot}.

\begin{table}[ht]
\begin{center}
\begin{tabular}{|c|c|c|}
\hline
Conformal Field &  ${SU^{*}(4)}_{D}$ & $USp(8)$
\\ \hline

$\phi^{[ABCD]|}(x)$  & (0,0,0)       & 42     \\ \hline
$\lambda_{\hat{\alpha}}^{[ABC]|} (x) $ & (1,0,0)
& 48\\ \hline

$h_{(\hat{\alpha}\hat{\beta})}^{[AB]|}(x)$ & (2,0,0)       & 27     \\ \hline
$\psi_{(\hat{\alpha}\hat{\beta}\hat{\gamma})}^A (x) $& (3,0,0) & 8 \\ \hline
$R_{(\hat{\alpha}\hat{\beta} \hat{\gamma}\hat{\delta})}(x) $& (4,0,0) & 1 \\ \hline
\end{tabular}
\end{center}
\caption {\label{USp(8)_doubleton} \small
Doubleton supermultiplet of $OSp(8^*|8)$ corresponding to the lowest weight vector
$|\Omega\rangle = |0\rangle$ with $SU^*(4)$ Dynkin labels and dimensions
of the $USp(8)$  representations.  The conformal fields are labelled by
$SU^*(4)$ spinor indices $\hat{\alpha}, \hat{\beta},\dots$ and $USp(8)$ indices $A,B,\dots$.}
\end{table}

\subsection{Constrained on-shell superfields for the six-dimensional $(4,0)$ supermultiplet
\label{6dsuperfield}}

The fields of the $(4,0)$ supermultiplet in six dimensions can be organized in an on-shell
superfield obeying two simple algebraic and differential constraints.
Following the conventions in \cite{Howe:1983fr} (see also \cite{fs,fs2}) the coordinates of the  six-dimensional
extended superspace are
\be 
\big( x^{\ha \hb}, \theta^{\hat \alpha}_A \big) \qquad \ha, \hb = 1, \dots, 4 \ ;
\quad A=1, \dots, 8 \ ;
\ee
as mentioned previously, $SU^*(4)$ spinor indices are denoted by hatted Greek
letters $\ha, \hb  \dots$; we will denote $USp(8)$ $R$-symmetry indices
by capital Latin letters $A,B,C \dots$.
The superspace covariant derivative is defined as
\be 
D^A_{\hat \alpha} = \partial^A_{\hat \alpha} + 
i \Omega^{AB} \theta^\hb_B \partial_{\hat \alpha  \hb} \ ,
\ee
where $\partial^A_{\hat \alpha} \theta^\hb_B = \delta^A_B \delta^\hb_\ha$. In the above expression $\Omega_{AB}=-\Omega_{BA}$
is a symplectic matrix which obeys
\be 
\Omega_{AB} \Omega^{BC} = \delta^C_A \ .
\ee
The symplectic matrix $\Omega_{AB}$ can be used to raise or lower indices as follows,
\be 
\theta^{A\ha} = \Omega^{AB} \theta^\ha_B, \qquad D_{A\ha}= \Omega_{AB} D^B_\ha \ .
\ee
With these definitions, the superspace derivatives obey the anticommutation relations
\be \{ D^A_\ha, D^B_\hb \} = 2 i \Omega^{AB} \partial_{\ha \hb} \ ,
\qquad \{ D_{\ha A}, D^B_{\hb} \} = 2 i \delta^B_A \partial_{\ha \hb} \ . \ee

To construct the superfield containing the fields of the $(4,0)$ multiplet we begin,
following the  analysis in \cite{Arvidsson:2003pc} of the $(2,0)$ multiplet,
with the superfield
\be
\Phi^{ABCD}\big( x^{\ha \hb}, \theta^\ha_A \big) 
\ee
transforming in the ${\bf 42}$ representation of $USp(8)$, {\it i.e.} its 
$USp(8)$ indices $A,B,C,D$ are antisymmetrized and obey a symplectic-traceless condition
\be
\Phi^{ABCD} \Omega_{CD}= 0\ .
\label{algebraicconstraint}
\ee
To have the right number of independent degrees of freedom we also need to impose a
differential constraint of the form
\be
D^A_\ha \Phi^{BCDE} + {1 \over 21} D_{\ha F} \Big\{ \Omega^{A[B} \Phi^{CDE]F}
+ {3 \over 4} \Omega^{[BC} \Phi^{DE]AF} \Big\}=0~~.
\label{diffconstraint}
 \ee
This differential constraint is analogous to the ones appearing in the superfield descriptions of the
$(2,0)$ supermultiplet in six dimensions and of the ${\cal N}=8$ supermultiplet in four dimensions
(Equation \ref{diffcon4D}).
The numerical coefficient ${1 \over 21}$ in the second term is chosen such that (\ref{diffconstraint}) sets to 
zero all $USp(8)$ 
representations appearing in the product ${\bf 8}\times {\bf 42}$ except for the 48-dimensional 
one. Thus, (\ref{diffconstraint}) eliminates all spin-${1 \over 2}$ fields except for those transforming in the ${\bf 48}$
representation of $USp(8)$.
Furthermore, we can define the additional superfields
\bea \Lambda_\ha^{ABC} = {2 \over 7} i D_{\ha D} \Phi^{DABC} & &
H_{\ha\hb}^{AB} = -{3 \over 16} D_{\ha C} \Lambda^{CAB}_\hb \no \ ,  \\
\Psi_{\ha \hb \hg}^{A} = {4 \over 27} i D_{\ha B} H^{BA}_{\hb \hg}  \ , & & 
R_{\ha \hb \hg \hd} = - {1\over 8} D_{\ha A} \Psi^{A}_{\hb \hg \hd} \ . 
\label{def-superfields}
\eea
They contain the same degrees of freedom as  $\Phi^{ABCD}$ except that their lowest components
are the component fields of the $(4,0)$ supermultiplet:
$\lambda^{ABC}_\ha , h^{AB}_{\ha \hb}, \psi^A_{\ha \hb \hg}$ and $R_{\ha \hb \hg \hd}$,
respectively\footnote{With potential abuse of notation, we will denote by $R$ both
the six-dimensional field transforming in the $(4,0,0)$ representation of $SU^*(4)$ and
the corresponding superfield. The field $R$ will be sometimes referred to as the \emph{generalized} graviton field strength 
since it reduces to the four-dimensional Riemann tensor under dimensional reduction}. 
It is not difficult to check that the superfields in equation~(\ref{def-superfields})
are symmetric in the spacetime indices.

The differential constraint (\ref{diffconstraint}) can be rewritten in terms of the superfields
introduced in equation~(\ref{def-superfields}),
\be
D^A_\ha \Phi^{BCDE} = -{i \over 6} \Big( \Omega^{A[B} \Lambda^{CDE]}_\ha + {3 \over 4}
\Omega^{[BC} \Lambda_\ha^{DE]A} \Big)  ~~.
\ee
This expression can be used to show that the superfields in (\ref{def-superfields}) should themselves
obey corresponding differential constraints,
\bea D_\ha^A \Lambda^{BCD}_\hb &=& 2 \partial_{\ha \hb} \Phi^{ABCD} 
- {1 \over 2} \Omega^{A[B} H^{CD]}_{\ha \hb}
+{1 \over 6} \Omega^{[BC} H^{D]A}_{\ha \hb} \label{ansatz-lambda} \no \ , \\
D^A_\ha H^{BC}_{\hb \hg} &=& i \partial_{\ha ( \hb} \Lambda^{ABC}_{ \hg)}
- i \Omega^{A[B} \Psi^{C]}_{\ha \hb \hg}
- {i \over 4}  \Omega^{BC} \Psi^A_{\ha \hb \hg} \no \ ,  \\
D^A_\ha \Psi^B_{\hb \hg \hd} &=&
{1 \over 3} \partial_{\ha ( \hb } H^{A B}_{\hg \hd)} - \Omega^{AB} R_{\ha \hb \hg \hd} 
\label{ansatz-Psi} \ , \no  \\
D^A_\ha R_{\hb \hg \hd \hat \epsilon} &=& 
{i \over 12} \partial_{\ha ( \hb} \Psi^A_{\hg \hd \hat \epsilon)} \ .
\label{ansatz-R}  \eea
The normalization constants in (\ref{def-superfields}) have been chosen to keep the numerical 
factors in the above expressions simple.
Note that these differential constraints do not impose any further constraint on the lowest
component fields $\lambda^{ABC}_\ha , h^{AB}_{\ha \hb}, \psi^A_{\ha \hb \hg}$ and
$R_{\ha \hb \hg \hd}$, which all correspond to independent degrees of freedom transforming
in the ${\bf 48}$, ${\bf 27}$, ${\bf 8}$ and singlet representations of $USp(8)$ respectively.
Moreover, if we expand the above differential constraints and  take the $\theta^0$ components,
we can read off the supersymmetry transformation rules of the fields of the $(4,0)$ supermultiplet,
\bea 
\delta_\eta \phi^{ABCD}_\ha &=&   {i \over 6} \eta^{\ha [A}  \lambda^{BCD]}_{\ha}
-{i \over 8} \eta^\ha_E \Omega^{[AB} \lambda^{CD]E}_{\ha} \label{transf-phi} \no \ , \\
\delta_\eta \lambda^{ABC}_\ha &=&  2 \eta^\hb_D \partial_{\ha \hb} \phi^{ABCD} 
+ {1 \over 2} \eta^{\hb [A}  h^{BC]}_{\ha \hb}
+{1 \over 6} \eta^\hb_D \Omega^{[AB} h^{C]D}_{\ha \hb} \label{transf-lambda} \no \ , \\
\delta_\eta h^{AB}_{\ha \hb} &=& i \eta^\hg_C \partial_{  \hg (\ha } \lambda^{ABC}_{\hb) } 
+ i \eta^{ \hg [A}  \psi^{B]}_{\ha \hb \hg}
- {i \over 4} \eta^\hg_C \Omega^{AB} \psi^C_{\ha \hb \hg}  \ , \no  \\
\delta_\eta \psi^A_{\ha \hb \hg } &=&
- {1 \over 3} \eta^\hd_B \partial_{\hd ( \ha } h^{A B}_{\hb \hg )} 
+ \eta^{ \hd A} R_{\ha \hb \hg \hd} \label{transf-Psi}  \ , \no  \\
\delta_\eta R_{\ha \hb \hg \hd } &=& 
{i \over 12} \eta^{\hat \epsilon}_A \partial_{\hat \epsilon ( \ha} \psi^A_{\hb \hg \hd )} \ .
\label{transf-R}  
\eea
In particular, the fact that these relations only involve the fields
$\phi^{ABCD}, \lambda^{ABC}_\ha , h^{AB}_{\ha \hb}, \psi^A_{\ha \hb \hg}
$ and $R_{\ha \hb \hg \hd}$ implies that these fields form a representation of the $OSp(8^*|8)$
superalgebra without having to add any extra degrees of freedom. This is in agreement with the 
oscillator analysis of the previous sections.


\subsection{Dimensional reduction of six-dimensional conformal  supermultiplets down to four dimensions}

We now consider the dimensional reduction of six-dimensional $(2,0)$ conformal doubleton supermultiplet of  \OS to four dimensions. It is described by the supermultiplet of coherent states
 \[ 
| \phi^{[AB]|}(x) \rangle \oplus
| \lambda_{\hat{\alpha}}^{A} (x) \rangle \oplus
| h_{(\hat{\alpha} \hat{\beta})}(x) \rangle  \ ,
\] 
where $A,B,\cdots$ denote $USp(4)$ indices and $\hat{\alpha}, \hat{\beta}, \cdots =1,2,3,4$
denote the spinor indices of $SU^*(4)$. The spinor representation of $SU^*(4)$ decomposes
with respect to the four-dimensional Lorentz group $SL(2,\mathbb{C})$ as
\[ {\bf 4} = \Big( {1 \over 2},0 \Big) \oplus \Big(0, {1 \over 2} \Big)~~; \]
this is the same as the decomposition of the fundamental representation of $SU(4)$ with respect to the compact subgroup  $SU(2)\times SU(2)$.
Thus we have the decomposition
\eq
 | h_{(\hat{\alpha} \hat{\beta})}(x) \rangle  = | h_{(\alpha \beta)}(x) \rangle  \oplus | h_{\alpha \dot{\beta}}(x)\rangle  \oplus | h_{(\dot{\alpha} \dot{\beta})}(x)\rangle \ ,
 \en
 where the coordinate vector $x_\mu$ is restricted to the four-dimensional Minkowski space.
Comparing with  the doubleton supermultiplet of $SU(2,2|4)$ \cite{GM1} we see that 
$| h_{(\alpha \beta)}(x) \rangle $  and $| h_{(\dot{\alpha} \dot{\beta})}(x)\rangle $ are simply the self-dual and anti-self-dual components of the field strength of a vector field in four dimensions.
The coherent state $| h_{(\alpha \dot{\beta})}(x)\rangle $ is the derivative of a scalar field
\eq
| h_{\alpha \dot{\beta}}(x)\rangle = \partial_{\alpha \dot{\beta}} | \phi^0(x) \rangle~~.
\en
The spinorial coherent state decomposes simply as
\eq
| \lambda_{\hat{\alpha}}^{A} (x) \rangle =| \lambda_{\alpha}^{A} (x) \rangle \oplus | \lambda_{\dot{\alpha}}^{A} (x) \rangle~~.
\en
The scalar coherent states $| \phi^{[AB]|}(x) \rangle  $  descend directly to four dimensions.
The resulting four-dimensional supermultiplet is in fact the ${\cal N}=4$ Yang-Mills 
supermultiplet with 6 scalar fields, 4 spin-${1 \over 2}$ fields and a vector field, and corresponds 
simply to  the doubleton (minimal) unitary supermultiplet of $PSU(2,2|4)$. The $R$-symmetry 
group $USp(4)$ of six-dimensional supermultiplet gets extended to $SU(4)$ under which 
$5+1$ scalars transform in the antisymmetric tensor representation. Four six-dimensional 
spinors of $SU^*(4)$ decompose into 4 chiral and 4 anti-chiral $SL(2,\mathbb{C})$ spinors 
in four dimensions, which also transform in ${\bf 4}$ and $\bar{ \bf 4}$ representations of the four-dimensional $R$-symmetry $SU(4)$.

\subsection{Dimensional reduction of six-dimensional  $(4,0)$  supermultiplet}

The fields of the $(4,0)$ doubleton supermultiplet of  the six-dimensional superconformal 
group $OSp(8^*|8)$ were listed in Table \ref{USp(8)_doubleton}. Let us now discuss the 
dimensional reduction of this conformal supermultiplet using coherent state formalism. 

The coherent state corresponding to the generalized graviton field strength decomposes as follows under dimensional 
reduction to four dimensions:
\be
|R_{(\hat{\alpha}\hat{\beta} \hat{\gamma}\hat{\delta})}(x)\rangle = |R_{(\alpha\beta\gamma\delta)}(x) \rangle \oplus |R_{(\dot{\alpha}\dot{\beta}\dot{\gamma}\dot{\delta})}(x) \rangle
\oplus  |R_{(\alpha\beta\gamma)\dot{\delta}}(x) \rangle \oplus |R_{(\dot{\alpha}\dot{\beta}\dot{\gamma})\delta}(x) \rangle \oplus 
|R_{(\alpha\beta)(\dot{\gamma}\dot{\delta})}(x) \rangle \ . \qquad  \no \ee
The  coherent states $|R_{(\dot{\alpha}\dot{\beta} \dot{\gamma}\dot{\delta})}(x)\rangle $ and 
$ |R_{(\alpha\beta\gamma\delta)}(x) \rangle$ correspond to the chiral and anti-chiral components of the field strength of 
the  four-dimensional graviton. 
\begin{table}[ht]
\begin{center}
\setlength{\extrarowheight}{2pt}
\begin{tabular}{| c | c | }
\hline
$6D$ Field & $4D$ Decomposition   \\
\hline
$R_{(\hat{\alpha}\hat{\beta} \hat{\gamma}\hat{\delta})}$ & 
$ R_{(\alpha\beta\gamma\delta)}  \oplus R_{(\dot{\alpha}\dot{\beta}\dot{\gamma}\dot{\delta})}
\oplus  \partial_{\dot{\delta}(\gamma} h^0_{\alpha\beta )}  \oplus 
\partial_{\delta (\dot{\gamma}} h^0_{\dot{\alpha}\dot{\beta} )}
\oplus \partial_{\alpha (\dot{\gamma}} \partial_{\dot{\delta})\beta} \phi^0$\\
\hline
$\psi^A_{(\ha\hb\hg)}$ & $\psi^A_{(\alpha\beta\gamma)}  \oplus
\psi^A_{(\da\db\dg) } \oplus \partial_{\dg(\alpha} \lambda^A_{\beta)} \oplus \partial_{\alpha(\db} \lambda^A_{\dg)}$  \\ 
\hline
$h^{[AB]|}_{\ha\hb}$ & $h^{[AB]|}_{\alpha\beta} \oplus h^{[AB]|}_{\da\db} \oplus \partial_{\alpha\db} \phi^{[AB]|}$ \\
\hline
$\lambda^{[ABC]|}_{\ha} $ & $\lambda^{[ABC]|}_{\alpha} \oplus \lambda^{[ABC]|}_{\da}$ \\
\hline
$\phi^{[ABCD]|}$ & $\phi^{[ABCD]|}$ \\
\hline
\end{tabular}
\caption{\small Reduction to four dimensions of the fields of the $(4,0)$ multiplet. \label{Tab4dreduction}}
\end{center}
\end{table} 

The  coherent states $ |R_{(\dot{\alpha}\dot{\beta}\dot{\gamma})\delta}(x) \rangle $ and
$ |R_{(\alpha\beta\gamma)\dot{\delta}}(x) \rangle $  are descendants 
of the coherent states corresponding to self-dual and anti-self-dual components of the field strength of a four-dimensional vector field.
The coherent state $|R_{(\alpha\beta)(\dot{\gamma}\dot{\delta})}(x) \rangle$ corresponds to the descendant of a scalar field.
The other fields are reduced to four dimension with the same procedure and listed in Table \ref{Tab4dreduction}. 
The primary conformal fields in four dimension coincide precisely with the fields of the CPT-self-conjugate doubleton supermultiplet constructed in \cite{GM2} which we reviewed in Section \ref{multipletN8}.

\subsection{ Gauge potentials of the $(4,0)$ supermultiplet}
\label{secpot}

The components of supermultiplets obtained through the oscillator method are the dynamical gauge invariant 
degrees of freedom of the underlying theory. In particular, the oscillator method  yields covariant field strengths 
and not gauge potentials.  

For example, the ${\cal N}=4$ super-Yang-Mills supermultiplet obtained through this method contains 
the states transforming in the $(1,0) \oplus (0,1)$ representation of the Lorentz group \cite{GM1}. 
In spinor notation, they are represented by two-index symmetric tensors , $F_{ \alpha  \beta}$ and 
$F_{\dot{\alpha}\dot{\beta}}$, each of them transforming irreducibly  the Lorentz group.
By contracting with the appropriate products of Pauli matrices they may be transformed to self-dual and 
anti-self-dual antisymmetric tensors carrying vector indices:
\eq
{1 \over 2 } \big( F_{\mu\nu} + * F_{\mu\nu} \big) =  F_{\alpha\beta} \sigma_{\mu\nu}^{\alpha\beta} ~ .
\en
Here $F_{\mu\nu}$ is the field strength of a Yang-Mills potential, 
$F_{\mu\nu} = \partial_\mu A_\nu -\partial_\nu A_\mu $ and $\sigma_{\mu\nu}$ is one of the diagonal 
$2\times 2$ blocks of Lorentz generators in the spinor representation. 

In this subsection we shall construct gauge potentials whose field strengths are the components of the 
$(4,0)$ supermultiplet given above. 
Since field strengths are obtained by taking derivatives of gauge potentials with respect to 
spacetime coordinates, it is standard practice  to convert the field strengths given 
in terms of spinorial indices into vectorial indices by  contraction with products of generalized Pauli
matrices in six dimensions ({\it i.e.} the off-diagonal blocks of the six-dimensional Dirac matrices in 
the Weyl representation). 
For the $(2,0)$ doubleton supermultiplet this conversion was given already in \cite{gnw}. 
For the symmetric tensor field $h_{(\alpha\beta)}$ one finds that it is associated to a two-form gauge 
field whose field strength is self-dual
\eq
h_{MNP} = \partial_{[M} b_{NP]} = * h_{MNP} \ .
\en
 Since a vector in six dimensions can also be described by a bispinor, 
one can represent the two form field $b_{MN}$ that transforms in the adjoint representation of 
$SU^*(4)$ by a field $ b^{\hat \alpha}_{\hat \beta}$ with one upper and one lower spinor index subject to 
$b^{\hat \alpha}_{\hat \alpha}=0$. Then the field strength is
\eq
h_{\hat \alpha \hat \beta} = h_{( \hat \alpha  \hat \beta)}= 
\partial^{\vphantom{\hat \gamma}}_{ \vphantom{\hb} \hat \gamma ( \ha} b^{ \hat \gamma}_{ \hat \beta)}
\en
and the gauge parameter is a vector $\chi^{[\ha\hb]} $ of $SU^*(4)$. 
The gauge potential for the generalized graviton field strength $R_{\ha \hb \hg \hd}$ is a tensor $C^{\hat \a }_{( \hb \hg \hd)}$, carrying fundamental indices of 
$SU^*(4)$, subject to the tracelessness condition
\eq
C^{ \hg}_{\ha \hb \hg } =0 \ . 
\en
The corresponding gauge transformation of this potential is
\be  C^{\ha}_{\hb \hg \hd} \rightarrow C^{\ha}_{\hb \hg \hd} + 
\partial_{ \hat \omega ( \hb }^{\vphantom{\ha}} \chi^{\hat \omega \ha}_{\hg \hd)} \ee
with the gauge parameter $\chi^{[\ha \hb]}_{(\hg \hd)}$ subject to the tracelessness condition,
\be 
\chi^{\ha \hg}_{\hb \hg} = 0 \ . 
\ee
Similarly, the gauge potential corresponding to the generalized gravitino field strength is a tensor
$\psi^{\ha}_{(\hb \hg)}$ which is also traceless, 
\be 
\psi^{\hb}_{\ha \hb}=0 
\ee
and transforms under a gauge transformation as
\be  
\psi^{\ha }_{\hb \hg}  \rightarrow \psi^{\ha}_{\hb \hg} + 
\partial_{\hat \omega  ( \hb } \chi^{\hat \omega \ha }_{\hg)} \ ;
\ee
the gauge parameter $\chi^{[\ha \hb]}_\hg$ satisfies the condition $\chi^{\ha \hb}_\hb =0$. 
It is important to note that the local gauge symmetry of the generalized gravitino field is distinct from 
supersymmetry,  as it can be inferred from the fact that the gauge parameter $\chi^{\ha \hb}_\hg$ 
does not transform in the fundamental representation of $SU^*(4)$.  

In Table  \ref{4_0_gaugepotentials}, we list the gauge fields corresponding to the $(4,0)$ supermultiplet and 
their gauge parameters in both vectorial as well as spinorial notation of the six-dimensional Lorentz 
group $SU^*(4)$. 
The gauge potentials in vector notation agree with the results of \cite{Hull:2000zn,Hull:2000rr}, 
with the exception of the potential $C$, which in our case is related to the generalized 
Riemann tensor by only one derivative.  

\begin{table}[t]
\begin{center}
 \begin{tabular}{|c@{ }|@{}c@{ }|@{}c@{ }|@{ }c@{ }|@{ }c@{ }|}
\hline
\textbf{\small $SO(4)$ rep} & \textbf{ \small Vectorial field} & \textbf{\small Constraints}  
& \textbf{\small $SU^*(4)$ rep} & \textbf{\small Spinorial field} \\
\hline \hline  
$(2,0)$ 
& $R_{[MNO][PQR]} 
$ & $ \! \begin{array}{c} *R=R* = R \\  R_{MNOPQR}=R_{PQRMNO} \end{array}   $  & $\fourbox$ & $R_{(\ha \hb \hg \hd)}=\partial^{\vphantom{ \hat \omega \hat \lambda}}_{ \hat \lambda ( \ha}  C^{\hat \lambda }_{\hb \hg \hd)}$ \\
 & $C_{[MNO][PQ]}$ & $\begin{array}{c} *C = C \\ C_{[MNOP]Q}=0  \end{array}$  & $\threeoneoneonebox$ & $C^{\ha}_{(\hb \hg \hd)}, 
\quad C^{\hg }_{\ha \hb \hg }=0$  \\
\hline
$({3\over 2},0)$ & $\psi_{[MNO]\ha} $ & $*\psi=\psi$ & $\threebox$ & $\psi_{(\ha \hb \hg)} = 
\partial^{\vphantom{\hat \lambda}}_{\hat \lambda ( \ha } \psi^{\hat \lambda}_{\hb \hg)}$ \\
 & $\psi_{[MN]\ha}$ & & $\threeoneonebox$ & $\psi^\ha_{(\hb \hg)}, \quad \psi^\hb_{\ha \hb} = 0$ \\
\hline
$(1,0) $ & $h_{[MNP]}$  & $*h=h$ &  $\twobox$ & $h_{(\ha \hb)}= \partial^{\vphantom{\hg}}_{\hg (\ha} b^{\hg}_{\hb)} $ \\
& $b_{[MN]}$ & & $\twooneonebox$ & $b^\ha_\hb, \quad b^\ha_\ha=0$ \\
\hline
$ ({1 \over 2},0)$ & $\lambda_\ha$ & & $\onebox$ & $\lambda_\ha$ \\
\hline 
$ (0,0)$ & $\phi$ & & $ 1  $ & $\phi$  \\ \hline \end{tabular} 
\end{center}
\caption{\label{4_0_gaugepotentials} \small
Correspondence between fields of the $(4,0)$ supermultiplet with vectorial indices and spinorial indices. 
Fields are organized according to the corresponding representations of the little group $SO(4)=SU(2)\times SU(2)$.
Lower and higher spinor indices correspond to 
fundamental and anti-fundamental $\mathbf{4}$ and $\overline{\mathbf{4}}$ indices respectively.} 
\end{table}

By examining the gauge potentials written in the vectorial notation, 
it is clear that the $(4,0)$ theory contains "higher-spin" gauge fields in the sense that 
the generalized graviton field $C_{MNO,PQ}$  has six-dimensional vector indices with mixed symmetry. 
Theories of this sort were studied in \cite{Labastida:1986gy,Campoleoni:2008jq}
and are the object of several no-go theorems in the literature. 
Constraints on an interacting $(4,0)$ theory from such higher-spin no-go theorems 
will be discussed in Section \ref{Secnogo}.

We have seen in the previous section that the fields of the $(4,0)$ supermultiplet
reduce to supermultiplets in five and four dimensions that do not have any 
fields with spin greater than two. In particular, in six dimensions we have
two different fields that reduce to the massless graviton in five and four dimensions, 
namely the generalized graviton $C$ and the metric graviton $g$ of Poincar\'e supergravity, 
just like the four-dimensional ${\cal N}=4$ Yang-Mills supermultiplet can be obtained from 
the conformal $(2,0)$ theory as well as from the non-conformal $(1,1)$ super-Yang-Mills 
theory in six dimensions. This is a special feature of the $(4,0)$ multiplet which is not present  
in a generic higher-spin theory in six dimensions.

\subsection{The $(4,0)$ supergravity multiplet as the square of the $(2,0)$ multiplet}

The degrees of freedom of the $(4,0)$ supermultiplet constructed in the previous section can be brought into direct correspondence with the states
obtained from the direct product of two six-dimensional $(2,0)$ multiplets.\footnote{This is similar to the fact that the fields in the tensor product of two $(1,1)$ vector multiplets are
in one to one correspondence with the fields of the $(2,2)$ Poincar\'e graviton multiplet.} 
In order to do so we need to
consider the transformation properties of the states of the $(4,0)$ multiplet under an
$USp(4) \times USp(4)$ subgroup of the $USp(8)$ $R$-symmetry group of the theory. 

The fundamental representation of $USp(8)$ decomposes as
\be 
{\bf 8} = ({\bf 4},{\bf 1}) \oplus ({\bf 1},{\bf 4}) \ .
\ee
We will denote as $a,b = 1, \dots, 4$ and $\check  a, \check  b=1, \dots, 4$ the
fundamental indices of the two $USp(4)$ groups, so that an $USp(8)$ index $A=1, \dots, 8$ 
is represented by the pair $A=(a, \check  a)$.
The $USp(8)$ symplectic matrix $\Omega$ can be brought to a block-diagonal form 
so that $\Omega_{a \check  a}=\Omega^{a \check  a}=0$.
The various component fields of the multiplet decompose in irreducible representations of $USp(4) \times USp(4)$.
This decomposition is listed in Table \ref{40-decompose}.
To obtain the $(4,0)$ multiplet, we now need to take the direct product of two copies of the $(2,0)$ multiplet
each transforming under a different $USp(4)$ subgroup of the $USp(8)$ $R$-symmetry group.
As we have seen, each $(2,0)$ multiplet has three component fields.

\begin{table}[t]
\begin{center}
\setlength{\extrarowheight}{2pt}
\begin{tabular}{| c | c | c |}
\hline
Field & Decomposition & $USp(4) \times USp(4)$ representations  \\
\hline
$\quad \phi^{[ABCD]|}$ & $ \phi^{[ab]|[\check  a \check  b ]|}   \oplus   \phi^{a \check  a}   \oplus  \phi $ &
$ ({\bf 5}, { \bf 5} ) \oplus  ({\bf 4},{\bf 4})  \oplus (1,1)$ \\[1pt] 
\hline
$ \lambda^{[ABC]|}_\ha $&  $ \lambda^{\check  a [ab]|}_\ha   \oplus  \lambda^{ a [ \check  a \check  b]|}_\ha
\oplus  \lambda^{a}_\ha   \oplus  \lambda^{\check  a}_\ha $ &
$ ({\bf 5}, { \bf 4} ) \oplus  ({\bf 4}, { \bf 5} ) \oplus ({\bf 4},1)  \oplus (1,{\bf 4})$ \\
\hline
$h^{[AB]|}_{(\ha \hb)}$ & $h^{[ab]|}_{(\ha \hb)} \oplus h^{[\check  a \check  b]|}_{(\ha \hb)} \oplus h^{a \check  a }_{(\ha \hb)} \oplus
h_{(\ha \hb)} $ & $( {\bf 5} ,1 ) \oplus ( 1, {\bf 5} ) \oplus ( {\bf 4} , {\bf 4} ) \oplus (1,1) $  \\
\hline 
$\psi^A_{(\ha \hb \hg)} $ & $\psi^a_{(\ha \hb \hg)} \oplus \psi^{\check  a}_{(\ha \hb \hg)}$ & $({\bf 4},1) \oplus (1, {\bf 4}) $\\
\hline
$R_{(\ha \hb \hg \hd)} $ & $ R_{(\ha \hb \hg \hd)} $ & $(1,1)$ \\
\hline
\end{tabular}
\medskip
\caption{\label{40-decompose} \small Decomposition of the component fields of the $(4,0)$ supergravity multiplet in irreducible
representations of the $USp(4) \times USp(4)$ subgroup of $USp(8)$.}
\setlength{\extrarowheight}{2pt}
\bigskip
\begin{tabular}{| c | c | c | c |}
\hline
  & $  h_{\hg \hd}$ & $\lambda^{\check  a}_\hg  $ & $ \phi^{[\check  a \check  b]|}$ \\[3pt]

\hline

$\;  h_{\ha \hb} $ & $\; R_{\ha \hb \hg \hd} \oplus \p_{\ha (\hg} h_{\hd) \hb} \oplus \p_{\ha ( \hg} \p_{\hd) \hb } \phi^0 \; $
&
$\quad \psi_{\ha \hb \hg}^{\check  a} \oplus \p_{\hg (\ha}^{\vphantom{{}_{|}}} \lambda^{\check  a}_{\hb)} \quad $ & $ \quad h^{[\check  a \check  b ]|}_{\ha \hb} \quad $ \\[3pt]
\hline
$\;  \lambda_\ha^a $ & $\psi^a_{\ha  \hg \hd } \oplus \p_{\ha ( \hg}^{\vphantom{{}_{|}}}  \lambda^a_{\hd)}$ & $h^{a \check  a}_{\ha \hg} \oplus \partial_{\ha \hg}\phi^{a \check  a}$
& $ \lambda^{a [\check  a \check  b]|}_\ha $ \\[3pt]
\hline
$\; \phi^{[ab]|} $ & $ h^{[ab]|}_{\hg \hd} $ & $\lambda^{\check  a [ab]|}_\hg $ & $ \quad \phi^{[ab]|[\check  a \check  b]|} \quad $ \\[3pt]
\hline
\end{tabular}
\medskip
\caption{\label{square20} \small The field content of the $(4,0)$ multiplet is obtained taking 
the direct product of two $(2,0)$ multiplets. Each one of the three rows and columns correspond 
to a different component field of the two $(2,0)$ multiplets. Each of the nine entries correspond 
to the decomposition of the product of two fields in irreducible representations of $SU^*(4)$ and 
$USp(4) \times USp(4)$. }
\end{center}
\end{table}
In each of the nine entries of Table \ref{square20}
we decompose a different product of two component fields in irreducible representations
of $SU^*(4)$. The states obtained in this way are labeled by $USp(4) \times USp(4)$ indices and exactly match the
 field content of the $(4,0)$ multiplet as listed in Table \ref{40-decompose}.\\

\renewcommand{\theequation}{5.\arabic{equation}}
\setcounter{equation}{0}

\section{Interacting  $(2,0)$ superconformal theory and $(4,0)$ theory}\label{interacting}

It is generally believed that $AdS/CFT$ duality between IIB superstring theory on 
$AdS_5\times S^5$ and ${\cal N}=4$ super Yang-Mills theory in four dimensions
has a higher-dimensional counterpart, relating M-theory on $AdS_7\times S^4$ and 
an interacting  $(2,0)$ superconformal theory in six dimensions.
The six-dimensional $(2,0)$ conformal supermultiplet, 
which was called the doubleton supermultiplet, appears as gauge modes in the 
compactification of 11-dimensional supergravity over $S^4$ and decouples from the Kaluza-Klein spectrum. 
However the entire Kaluza-Klein tower of 11 dimensional supergravity over $S^4$ can be obtained by 
tensoring of the $(2,0)$ doubleton supermultiplets and fitted into massless and massive supermultiplets of 
$OSp(8^*|4)$ \cite{gnw}. The $(2,0)$ supermultiplet  does not have a Poincar\'e limit as a representation 
of the $AdS_7$ supergroup $OSp(8^*|4)$ and it was pointed out in \cite{gnw} that the theory based on the 
$(2,0)$ doubleton supermultiplet must be a superconformal field theory that lives 
on the boundary of $AdS_7$ which is the six-dimensional Minkowski space. 
The $(2,0)$ superconformal multiplet consists of 5 scalars, four symplectic  Majorana-Weyl spinors, and one 
two-form gauge field   
whose field strength is  subject to six-dimensional self-duality. Due to self-duality, a local covariant free action for the 
 two-form gauge field vanishes identically. Hence one can have only covariant equations of motion for the 
 free $(2,0)$ theory. 

Using M/superstring theory arguments, Witten showed that there must exist interacting 
 six-dimensional superconformal theories
 based on the $(2,0)$ doubleton supermultiplet \cite{Witten:1995zh}. 
These theories are  classified by the Dynkin diagrams of  simply laced groups (A-D-E) 
and describe the low energy decoupling limits of IIB superstrings on a $K3$ manifold. 
Among these theories those belonging to the A-series also have M-theory description in terms 
of $M5$ branes \cite{Witten:1995zh}. Witten has argued that these interacting superconformal $(2,0)$
 theories may exist only as quantum theories without a classical description in terms of invariant
actions and/or covariant  equations of motion \cite{Witten:1997kz}. The reduction of the $(2,0)$ theory on 
a two torus yields the ${\cal N}=4$ super Yang-Mills theory in four dimensions.
Reducing the interacting $(2,0)$  theory on Riemann surfaces with punctures leads to a plethora of 
${\cal N}=2$ superconformal field theories in four dimensions \cite{Gaiotto:2009we} whose 
five-dimensional supergravity duals were studied in 
\cite{Gaiotto:2009gz}. These recent results lend further support for the existence of interacting 
(2,0) superconformal theories in six dimensions. 

The $(4,0)$ doubleton supermultiplet of $OSp(8^*|8)$ we studied above can be decomposed into 
supermultiplets of $OSp(8^*|4)$ subalgebra. Denoting the $USp(4)$ indices as $a,b,c=1,2,3,4$ we 
find that it contains one generalized graviton field strength $(2,0)$ supermultiplet consisting of the fields 
\eq
R_{(\hat{\alpha}\hat{\beta} \hat{\gamma}\hat{\delta})}(x),  \quad  \psi_{(\hat{\alpha}\hat{\beta}\hat{\gamma})}^a (x),  \quad  h_{(\hat{\alpha}\hat{\beta})}^{[ab]|}(x) \oplus h_{(\hat{\alpha}\hat{\beta})}(x) , \quad \lambda_{\hat{\alpha}}^{[abc]|} (x) , \quad \phi(x) 
\label{confgrav6d}\en
four $(2,0)$  generalized gravitino supermultiplets consisting of the fields 
\eq
 \psi_{(\hat{\alpha}\hat{\beta}\hat{\gamma})} (x),  \quad  h_{(\hat{\alpha}\hat{\beta})}^a(x) , \quad \lambda_{\hat{\alpha}}^{[ab]|} (x) \oplus \lambda_{\hat{\alpha}} (x) , \quad \phi^{[abc]|}(x)
\en
and five  $(2,0)$ doubleton supermultiplets containing the fields
\eq
  h_{(\hat{\alpha}\hat{\beta})}(x) , \quad \lambda_{\hat{\alpha}}^{a} (x)  , \quad \phi^{[ab]|}(x)
\en
Hence we expect any interacting theory of $(4,0)$ supermultiplet  to have a consistent truncation to an interacting  theory describing the coupling of a generalized $(2,0)$ graviton supermultiplet to five  
$(2,0)$ doubleton supermultiplets. If this interacting theory admits a limit where the $(2,0)$ graviton 
supermultiplet decouples one is then left with a theory of five $(2,0)$ doubleton supermultiplets. 

The corresponding decomposition of the four dimensional doubleton  supermultiplet of $SU(2,2|8)$ with respect to 
$PSU(2,2|4)$ leads to  ${\cal N}=4$ Weyl multiplet plus its CPT conjugate, four chiral ${\cal N}=4$ 
gravitino supermultiplets and their  CPT conjugates, and six Yang-Mills supermultiplets.
There exists twistor string theories in four dimensions \cite{Witten:2003nn, Berkovits:2004hg} which 
are invariant under $PSU(2,2|4)$ and, at least in principle, can be used to calculate scattering amplitudes 
of ${ \cal N}=4$ super Yang-Mills theory. 
One finds that twistor string description requires the introduction of ${\cal N}=4$ graviton supermultiplets 
coupled conformally to Yang-Mills supermultiplets \cite{Berkovits:2004jj}. The helicity of the conformal graviton 
that appears in scattering amplitude calculations is correlated with the helicity of the Yang-Mills gluons 
\cite{Brodel:2009ep}. Since  ${\cal N}=4$ super-Yang-Mills theory is parity invariant, both the positive and 
negative helicity conformal graviton supermultiplets must be present in the resulting theory. 

Since supertwistors exist in six dimensions it is natural to expects that there should exist a six-dimensional analog 
of either Witten's or Berkovits' twistor string theory, which can be used to calculate amplitudes of  an interacting
$(2,0)$ "gauge" theory and, similarly to their four-dimensional analogs, 
of the six-dimensional $(2,0)$ conformal graviton supermultiplets. 
The unique candidate for such a conformal graviton supermultiplet is the one given in eq.~(\ref{confgrav6d}).
Under reduction to four dimensions one then obtains  a CPT invariant theory describing the coupling of 
${\cal N}=4$ super-Yang-Mills theory to conformal supergravity.

\subsection{Perspectives from Weinberg and Coleman-Mandula theorems}\label{Secnogo}

As briefly mentioned in  Section~\ref{secpot}, the gauge potentials in the $(4,0)$ supermultiplet 
can be regarded as massless higher-spin fields with space-time indices of mixed symmetry.
Such theories have been studied at length in the literature and are expected to obey constraints 
following from higher-spin gauge invariance. 
In asymptotically flat space-times, several no-go theorems -- most notably Weinberg's theorem 
\cite{Weinberg:1964ew} and the  Coleman-Mandula theorem \cite{Coleman:1967ad} --
rule out possible interactions between fields of higher and lower spin under certain assumptions.  

The original Weinberg theorem showed that particles with spin greater than two cannot mediate 
long-distance interactions in a flat four-dimensional space-time \cite{Weinberg:1964ew}.
Specifically, the theorem considers the scattering of one higher-spin field $\varphi_{\mu_1 \dots \mu_S}(q)$ 
with $n-1$ scalars. In the \emph{soft} limit, in which the momentum $q$ of  the higher-spin field is small, 
the scattering amplitude behaves as
\be A_n (\phi_1, \dots, \phi_{n-1}, \varphi_{\mu_1 \dots \mu_S}) =
 A_{n-1}(\phi_1, \dots, \phi_{n-1}) \ \sum^{n-1}_{i=1}  g_i \ { p_i^{\mu_1} \dots  p_i^{\mu_S} \ \epsilon_{\mu_i \dots \mu_S}(q)
 \over 2 p_i \cdot q } \ , \label{Weinberg1} \ee
where $p_i$ are the momenta of the scalars fields, $\epsilon_{\mu_1 \dots \mu_S}$ is the polarization 
tensor for the higher-spin field and $A_{n-1}$ is a scattering amplitude involving only $n-1$ scalars.
An higher-spin gauge transformation acts on the polarization tensor as
\be 
\epsilon(q) \rightarrow \epsilon(q) + i q \Lambda(q) \ .  
\ee 
If there are no extra factors of $q$ present in the cubic vertices, 
a gauge transformation of the external higher-spin field could change the amplitude by a finite 
amount due to a cancellation of the poles in $p_i \cdot q$. Hence, in order to decouple the 
unphysical states and preserve gauge invariance in this limit, one needs to impose an 
extra condition on the momenta to cancel these contributions:
\be 
\sum^{n-1}_{i=1} g_i \ p_{i, \mu_1} \dots p_{i, \mu_{S-1}} = 0  \ ,  
\ee
where $g_i$ are the coupling constants corresponding to the $\varphi \phi \phi$ vertices. 
Unless $S=1$ or $S=2$, 
this condition is too restrictive and the $g_i$ are forced to vanish altogether.
The theorem can be further generalized to consider the scattering of fields of different spin 
\cite{Taronna:2010qq}. In this case, 
the problem arises only when the field exchanged has spin lower than the external higher-spin field. 
It is important to point out that 
Weinberg's theorem does not rule out all interactions in a theory with higher-spin fields. The argument applies only to amplitudes with 
 a single massless higher-spin field, assumes that the theory is local and restricts 
only the minimal cubic couplings of the higher-spin field with the other fields of the theory 
\cite{Sagnotti:2010at,Taronna:2010qq}.  
Extensions of Weinberg's theorem by Witten \cite{Weinberg:1980kq} 
and, more recently, by Porrati \cite{Porrati:2008rm} also
imply that higher-spin fields cannot be minimally coupled  to gravity in a consistent manner in flat spacetimes.

Coleman-Mandula theorem \cite{Coleman:1967ad} and its extension by Haag, Lopuszanski and Sohnius
\cite{Haag:1974qh} state that the maximal extension of Poincar\'e 
symmetry algebra is the semi-direct sum of a Poincar\'e (or conformal) superalgebra and a purely internal 
Lie algebra whose generators commute with those of the Poincar\'e (or superconformal) algebra. 
The main assumptions of Haag, Lopuszanski and Sohnius
 are that the 
S-matrix for elastic two-body scattering is nontrivial at all angles and 
that there are finitely many particles types below any given energy.

Despite the no-go theorems in the literature, in recent years many novel results were obtained 
showing that it is possible to have consistent interactions of higher-spin fields with lower spin-fields 
(including gravity) if one relaxes some of the assumptions.    
Interacting higher-spin theories in asymptotically flat spacetimes were successfully 
constructed using a lightcone gauge approach \cite{Bengtsson:1983pg,Bengtsson:1983pd,Metsaev:1993mj,Metsaev:2007rn} 
or techniques based on Noether's procedure \cite{Berends:1984rq,Manvelyan:2010jr,Ruehl:2011tk}
(see \cite{Bekaert:2010hw} for a comprehensive review).

In case of a putative interacting theory of the $(4,0)$ supermultiplet, 
both the higher-spin no-go theorems and the examples of higher-spin theories in the literature 
give interesting hints on the nature of the allowed interactions. 
Weinberg's theorem appears to exclude minimal cubic vertices involving the higher-spin field of the 
theory - the generalized graviton $C^{\alpha}_{ \b \gamma \delta}$  - 
and two fields of lower spin. 
In fact, contributions from such vertices in the soft-limit $q \rightarrow 0$ are 
already excluded by the mixed symmetry of the higher-spin field's polarization 
tensor in (\ref{Weinberg1}), which has at least two antisymmetric Lorentz indices. 
  
The fact that many known examples of consistent higher-spin theories involve non-local interactions
suggests that the interactions of the $(4,0)$ theory may also be non-local. 
Interestingly, interactions of this sort may be likely to appear in the $(2,0)$ theory.
The authors of \cite{Huang:2010rn, Czech:2011dk} have looked for possible three-point and four-points 
amplitudes in the $(2,0)$ theory which were rational in the kinematic invariants $s_{ij}=-(k_i+k_j)^2$ and 
found that, to write down interactions consistent with the superconformal symmetry, one may need to give 
up either six-dimensional Lorentz covariance \cite{Czech:2011dk} or to consider amplitudes which are not 
rational in $s_{ij}$.         

At this stage, further study is required to determine whether it is possible 
to introduce consistent interactions in a theory of the $(4,0)$ multiplet. 
In particular, it would be interesting to study 
candidate cubic vertices in the light-cone formalism of \cite{Metsaev:1993mj} or using the 
techniques of \cite{Manvelyan:2010jr,Ruehl:2011tk}. 
As mentioned in the previous section, the study of six-dimensional twistor 
strings could also offer some insight on an interacting $(4,0)$ theory.



\renewcommand{\theequation}{6.\arabic{equation}}
\setcounter{equation}{0}

\section{Discussion}
Maximal ungauged Poincar\'e  supergravity in six dimensions has $(2,2)$ supersymmetry  
with $R$-symmetry group  $USp(4)\times USp(4)$.  
Its U-duality group is $SO(5,5)$ with maximal compact subgroup $USp(4)\times USp(4)$
\cite{Tanii:1984zk}. 
The theory contains 5 antisymmetric tensor fields and 16 vector fields. The field strengths of the tensor fields 
together with their duals form a 10-dimensional representation of the $U$-duality group 
$SO(5,5)$, and the vector fields transform in the spinor representation. 

The equations of motion of linearized maximal supergravity in six dimensions are not 
conformally invariant. This is best seen by the fact that the $(2,2)$ graviton supermultiplet 
contains field strengths whose 
transformation properties under the Lorentz group $SU^*(4)$ 
are described by Young tableaux which have more than one row.
In contrast, field strengths of massless conformal fields in six dimensions have $SU^*(4)$ tableaux that contain only one row.

For the same reason,  the maximally $(1,1)$ supersymmetric Yang-Mills
theory cannot be conformally invariant in six dimensions.  Nevertheless, the six-dimensional  
$(1,1)$ supersymmetric Yang-Mills theory reduces to ${\cal N}=4$ super Yang-Mills theory  
in four dimensions which is conformally invariant, not only classically but also at the quantum level.

At the same time, strong evidence suggests that there exists a conformally invariant interacting
theory with fields in the $(2,0)$ doubleton supermultiplet of $OSp(8^*|4)$ that is dual to $M$-theory
on $AdS_7\times S^4$. It has been argued that 
this theory reduces to ${\cal N}=4$ super Yang-Mills in four dimensions as well
\cite{Witten:1995zh,Strominger:1995ac,Douglas:2010iu,Lambert:2010iw}.

This raises the question whether maximal  (ungauged) supergravity theories in six 
and four dimensions follow a similar pattern. That is,  since the $(2,2)$ six-dimensional Poincar\'e 
supergravity reduces to the ${\cal N}=8$ supergravity in four dimensions and the linearized field 
equations of the latter are conformally invariant, it is interesting to ask whether there exists a 
maximal supergravity in six dimensions whose field equations at the linearized level are conformally 
invariant and whose fields can be identified with  a conformal supermultiplet in six dimensions.
The theory of the $(4,0)$ supermultiplet we constructed in this paper is indeed a likely candidate 
since this multiplet is a representation of the six-dimensional superconformal algebra and it reduces to 
the maximal supergravity multiplet in four dimensions. 
Moreover, in close similarity with the four-dimensional maximal gauge and supergravity multiplets,
the six-dimensional $(4,0)$ and $(2,0)$ multiplets enjoy a KLT-like relation, the former being 
the tensor product of two copies of the latter. 
This $(4,0)$ supermultiplet would be the gravitational analog of the $(2,0)$ multiplet and, 
if an interacting non-metric theory based on it exists, then that theory   may
 potentially be equivalent, at the linearized level, to the $(4,0)$ conformal gravity theory studied in 
\cite{Hull:2000zn,Hull:2000rr} with a different formalism. 
We should perhaps stress once more that the fundamental 
  principles underlying the interacting $(2,0)$ theory are not yet known. 
Once these principles are uncovered we 
may be able to investigate the question whether the same principles can be extended to an interacting $(4,0)$ theory.

In the formalism of \cite{Hull:2000zn,Hull:2000rr},
field strengths and gauge potentials of the $(4,0)$ supermultiplet 
are written with vector indices of mixed symmetry. If the linearized multiplet can be promoted to a fully 
interacting theory, we expect higher-spin no-go theorems present in the literature to restrict the
allowed interactions. 
In particular, we would expect the mixed-symmetry field $C$ to be coupled non-minimally to the fields of lower spin. 

Remarkably, the $SU^*(4)$ representations of  the fields of six-dimensional $(4,0)$ 
supermultiplet remain irreducible with respect to the covering  group, 
$USp(2,2)$, of the five-dimensional Lorentz group $SO(4,1)$. Hence, under dimensional 
reduction to five dimensions the field content remains the same and, when restricted to 
the five-dimensional Lorentz group,  the  field content coincides with that of maximal 
supergravity in five dimensions. This is also consistent with the fact that $R$-symmetry 
group of maximal supergravity in five-dimensional is $USp(8)$. 

We do not expect that a non-metric maximal supergravity based on the $(4,0)$ supermultiplet admits a 
covariant action in six dimensions.  However, as suggested for the six-dimensional 
interacting $(2,0)$  theory, one may try to gain some insights into the six-dimensional theory by 
studying its dimensional reduction to five-dimensions \cite{Douglas:2010iu}.
Since the $(4,0)$ supermultiplet reduces to the fields of ${\cal N}=8$ Poincar\'e supergravity, 
we expect the 42 scalars to parameterize the coset space
\[
\frac{ E_{6(6)}}{USp(8)} 
\]
and the equations of motion of the presumed six-dimensional theory to 
have $E_{6(6)_{global}}\times USp(8)_{local}$ symmetry.
The self-dual tensor fields of the $(4,0)$ supermultiplet should then transform linearly 
in the ${\bf 27}$ of $E_{6(6)}$.

The conformal superalgebras discussed in this paper are valuable tools in the study of 
possible counterterms for maximal supergravity in various dimensions. In oscillator language, 
linearized supersymmetric counterterms can be obtained acting with fermionic generators 
on a suitable lowest weight state. 
In general, states corresponding to half-BPS linearized counterterms with four fields can 
be obtained by starting with four copies of the Fock vacuum and acting with all sixteen grade $+1$ 
fermionic generators. Similarly, candidate non-BPS counterterms can be obtained acting with 
32 odd generators on a given lowest-weight state, which needs to be a Lorentz scalar and
a singlet under the $R$-symmetry group.

In four dimensions we obtain the half-BPS state 
\be 4D: 
\qquad \Big( L^{i x} \Big)^{16} | 0,0,0,0 \rangle \ ,
\ee 
where the sixteen odd generators $L^{i x}$ are defined in Section~\ref{multipletN8}. 
This state is manifestly invariant under 32 supersymmetry generators and, upon acting with the intertwiner operator,
leads to the supersymmetric completion of the $R^4$ term which plays an important role in the 
analysis of candidate counterterms in four dimensions \cite{Beisert:2010jx,Elvang:2010jv,Elvang:2010xn,Kallosh:2011qt,Freedman:2011uc,Bossard:2011tq}. 
Note that the action of the $16$ generators above is equivalent to a superspace integral over half 
of the fermionic coordinates. 

Similarly, in three dimensions one can obtain a simple half-BPS term with four fields from the state 
\be 3D: \qquad \Big( A^{i \kappa} \Big)^{16} | 0,0,0,0 \rangle \, , \ee
where $A^{i \kappa}$ is defined in Section~\ref{Osposcillators3d}. Three-dimensional ${\cal N}=16$ supergravity has a better UV behavior than four-dimensional ${\cal N}=8$ supergravity, 
and it is expected to be finite if the four-dimensional theory is finite. 
It would be interesting to see if invariance 
under $E_{8(8)}$ and $SO(16)$ pose constraints 
on candidate counterterms in three dimensions which are more stringent than the ones 
from $E_{7(7)}$ and $SU(8)$ in four dimensions. 

In six dimensions it is possible to get a counterterm with four fields from the  state
\bea 
6D: \qquad \Big( A^{i x} \Big)^{16} | 0,0,0,0 \rangle  \ ,
\eea  
where $A^{i x} $ is defined in Section \ref{Osposcillators6d}. 
The four-graviton part of the six-dimensional counterterm contains four copies of the Weyl 
tensor $R_{\ha\hb\hg\hd}$ which belongs to the $(4,0)$ conformal supermultiplet, and can be written as,
\eq
6D: \quad R^4 = \epsilon^{\ha_1\hb_1\hg_1\hd_1} \epsilon^{\ha_2\hb_2\hg_2\hd_2} \epsilon^{\ha_3\hb_3\hg_3\hd_3}\epsilon^{\ha_4\hb_4\hg_4\hd_4} R_{\ha_1\ha_2\ha_3\ha_4}  R_{\hb_1\hb_2\hb_3\hb_4} R_{\hg_1\hg_2\hg_3\hg_4} R_{\hd_1\hd_2\hd_3\hd_4} \ .
\label{ctdeq6}
\en
The symmetries of $R_{\ha\hb\hg\hd}$ forbid other index contractions, suggesting that the 
structure in equation~(\ref{ctdeq6}) is unique.  Under dimensional reduction to four dimensions 
it reduces to the standard CPT-invariant $R^4$ term of the form
\be 
4D: \qquad R_{\alpha\beta\gamma\delta} 
                           R^{\alpha\beta\gamma\delta} 
                           R_{\da\db\dg\dd}R^{\da\db\dg\dd} \ .
\ee
The uniqueness of the $R^4$ term in four dimensions lends further support to the 
expectation that the corresponding $R^4$ term in six dimension (\ref{ctdeq6}) is unique as well.
Future work will investigate in detail the application of oscillator techniques to the study of BPS and 
non-BPS counterterms for maximal supergravity in various dimensions.\\

\vskip 0.3in

\noindent{\large \bf Acknowledgements}

\smallskip
We would like to thank H. Elvang, Y.-t. Huang, C. Hull and H. Samtleben for discussions  and  the Kavli Institute for Theoretical Physics for hospitality during
the course of this work. This research was supported in part by the National Science Foundation
under grant PHY-08-55356 and in part by the National Science Foundation
under grant PHY-05-51164. RR also acknowledges the support of the A.P.~Sloan Foundation.

\bigskip 

\bigskip

\appendix

\def\theequation{A.\arabic{equation}}
\setcounter{equation}{0}
\section{A Review of the Oscillator Construction of the Positive Energy Unitary Representations of the $4D$ Conformal Group $SU(2,2)$
\label{AppA}}

The generators of the conformal group in four dimensions $SU(2,2)$ (the two sheeted
covering of $SO(4,2)$) satisfy the commutation relations
\begin{eqnarray}
[M_{\mu\nu},M_{\rho\sigma}] & = &i (\eta_{\nu\rho}M_{\mu\sigma}-
\eta_{\mu\rho}M_{\nu\sigma} -\eta_{\nu\sigma}M_{\mu\rho}+
\eta_{\mu\sigma}M_{\nu\rho})\cr
[ P_{\mu}, M_{\rho\sigma} ] & = & i (\eta_{\mu\rho}P_{\sigma}-\eta_{\mu\sigma}
P_{\rho})\cr
[K_{\mu},M_{\rho\sigma}]& = &i (\eta_{\mu\rho}K_{\sigma}-\eta_{\mu\sigma}
K_{\rho})\cr
[D,M_{\mu\nu}]& = & [P_{\mu},P_{\nu}] = [K_{\mu},K_{\nu}]=0\cr
[P_{\mu},D] & = &iP_{\mu}; \quad [K_{\mu},D]=-iK_{\mu}\cr
[P_{\mu},K_{\nu}]& = &2i(\eta_{\mu\nu}D-M_{\mu\nu})
\end{eqnarray}
where $M_{\mu\nu}$ are the  Lorentz group
generators,  $P_{\mu}$ the four-momentum generators, $D$ the dilatation
generator and   generators
of  $K_{\mu}$  the special conformal generators.   The Lorentz metric is  $\eta_{\mu\nu}=\textrm{diag}(+,-,-,-)$ with ($\mu, \nu,
\dots = 0, 1, 2, 3$)

Defining
\eq
M_{\mu 5} = {1 \over 2} (P_{\mu} - K_{\mu}), \quad
M_{\mu 6} = {1 \over 2} (P_{\mu} + K_{\mu}), \quad M_{56} = -D,
\en
 the Lie algebra of  the conformal group can be written in a manifestly   $SO(4,2)$ covariant form
($-\eta_{55}=\eta_{66}=1$; $\quad a,b,\dots=0,1,2,3,5,6$)
\eq
[M_{ab}, M_{cd}] = i(\eta_{bc}M_{ad} - \eta_{ac}M_{bd}
-\eta_{bd}M_{ac} + \eta_{ad}M_{bc}).\label{SO42app}
\en
where $-\eta_{55}=\eta_{66}=1$; $\quad a,b,\dots=0,1,2,3,5,6$.
  Spinor representation of $SU(2,2)$ can be written in terms of four-dimensional gamma matrices $\gamma_{\mu}$ that satisfy
($\{ \gamma_{\mu},\gamma_{\nu}\}=2\eta_{\mu\nu}$) and
$\gamma_{5}=\gamma_{0}\gamma_{1}\gamma_{2}\gamma_{3}$ as follows:
\eqn
\Sigma_{\mu\nu} := \frac{i}{4}\left[\gamma_{\mu},\gamma_{\nu}\right]
\qquad
\Sigma_{\mu 5} :=  \frac{i}{2}\gamma_{\mu}\gamma_{5}
\qquad
\Sigma_{\mu 6} :=  \frac{1}{2}\gamma_{\mu}
\qquad
\Sigma_{56} :=  \frac{1}{2}\gamma_{5} \ .
\enn
%
%
We shall adopt the gamma matrix conventions of \cite{GMZ2}
\eqn
\gamma^{0}=\gamma_{0}=\left(\begin{array}{cc}
\mathbf{1} & 0  \\
0 & -\mathbf{1}\end{array}\right) \qquad
\gamma^{m}=-\gamma_{m}=\left(\begin{array}{cc}
0 & \sigma^{m}\\
-\sigma^{m} & 0\end{array}\right) \quad
\Rightarrow \quad \gamma_{5} =
i \left(\begin{array}{cc}
0 & \mathbf{1}  \\
\mathbf{1} &0 \end{array}\right)
\enn
where $\sigma^m, \, \, m=1,2,3 $ are the usual Pauli matrices .

The oscillator construction of the positive energy unitary representations  of the conformal group is simplest in the compact basis that is covariant with respect to its maximal compact subgroup, which is manifestly unitary.
The maximal compact subgroup $SU(2)_{L}\times SU(2)_{R}\times U(1)_{E}$
is generated by  the following generators
\eqn
L_{m}&=&\frac{1}{2} \left( \frac{1}{2} \varepsilon_{mnl} M_{nl}+M_{m5}\right)
\qquad
\longrightarrow SU(2)_{L}\nn\cr
R_{m}&=&\frac{1}{2} \left( \frac{1}{2} \varepsilon_{mnl} M_{nl}-M_{m5}\right)
\qquad
\longrightarrow SU(2)_{R}
\enn
where $m,n=1,2,3$. They
satisfy
\eqn
[L_{m},L_{n}]&=&i \varepsilon_{mnl}L_{l}\nn\cr
[R_{m},R_{n}]&=&i \varepsilon_{mnl}R_{l}\nn\cr
[L_{m},R_{n}]&=& [E,L_{n}]=[E,R_{n}] =0.
\enn
The $U(1)_{E}$
generator $E=\frac{1}{2}(P_{0}+K_{0})$  determines the three-grading and is  simply the
conformal Hamiltonian, whose spectrum is bounded from below in a positive energy representation.
Denoting the maximal compact Lie subalgebra of $SU(2)_{L}\times SU(2)_{R}\times U(1)_{E}$
as $\mathcal{L}^{0}$, the conformal algebra $g$ has a
three-graded decomposition
\eq
g = \mathcal{L}^{+} \oplus \mathcal{L}^{0} \oplus \mathcal{L}^{-},
\en
where
\eqn
[\mathcal{L}^{0},\mathcal{L}^{\pm}] =  \mathcal{L}^{\pm}, \qquad
[E,\mathcal{L}^{\pm}]&=&\pm \mathcal{L}^{\pm}
\enn
and the other commutators are equal to zero.

To construct positive energy unitary representations in a manifestly unitary way it is most convenient to work in the compact
$SU(2)_{L}\times SU(2)_{R}\times U(1)_{E}$ covariant basis.
The  positive energy lowest weight UIRs of
$SU(2,2)$
can then be constructed very simply  by using the oscillator method of
\cite{mgcs,ibmg,GM1,GMZ1}.

Consider now $P$ copies (``generations'' or "colors") of bosonic oscillators
 $a^{i}(R)=
a_{i}(R)^{\dagger}$, $b^{j}(R)=
b_{j}(R)^{\dagger}$ that satisfy
\eq
[a_i(R), a^j(S)] = \delta_{i}^{j} \delta_{RS}, \quad
[b_i(R), b^j(S)] = \delta_{i}^{j} \delta_{RS}
\en
with $i,j=1,2$; $R,S= 1, \dots, P$,
We form $P$ sets of spinors  $\Psi (R)$
\eq
\Psi(R) := \left(\begin{array}{c}
a_{1}(R)\cr
a_{2}(R)\cr
b^{1}(R)\cr
b^{2}(R)
\end{array}\right)
\en
whose Dirac conjugates are
\eq
\bar{\Psi}(R)\equiv {\Psi}^{\dagger}(R)\gamma^{0}=\left(a^{1}(R),a^{2}(R),
-b_{1}(R),- b_{2}(R)\right).
\en
Then the bilinears
\eq
\bar{\boldsymbol{\Psi}}\Sigma(M_{ab}) \boldsymbol{\Psi} :=
\sum_{R=1}^{P} \bar{\Psi}(R)\Sigma(M_{ab})\Psi(R),
\en
realize the Lie algebra of $SU(2,2)$
\eq
\left[ \bar{\boldsymbol{\Psi}}\Sigma(M_{ab})\boldsymbol{\Psi},\bar{\boldsymbol{\Psi}}\Sigma(M_{cd})\boldsymbol{\Psi}\right]=
\bar{\boldsymbol{\Psi}}\left[ \Sigma(M_{ab}),\Sigma(M_{cd})\right]\boldsymbol{\Psi},
\en
The Fock space of the bosonic oscillators decompose into an infinite number of  irreducible positive energy unitary representations of  $SU(2,2)$ that belong to the holomorphic discrete series.
In the above realization of $SU(2,2)$  non-compact generators are
given by linear combinations of di-creation and di-annihilation
operators of the form ${\mathbf{a}}^{i}\cdot {\mathbf{b}}^{j}$ and
${\mathbf{a}}_{i}\cdot {\mathbf{b}}_{j}$ \footnote{ The dot
product $\cdot$  denotes summation over the generation or color index $R$, {\it i.e.}
${\mathbf{a}}^{i}\cdot {\mathbf{b}}^{j}\equiv \sum_{R=1}^{P} a^{i}(R)b^{j}(R)$,
etc.}.
The  generators,
 $L_{m}$,  of $SU(2)_{L}$
 \eq
L^{k}_{i} := {\mathbf{a}}^{k} \cdot {\mathbf{a}}_{i}
-{1 \over 2} \delta^{k}_{i}N_{a},
\en
whereas the generators $R_{m}$ of $SU(2)_{R}$ are
\eq
R^{i}_{j} := {\mathbf{b}}^{i} \cdot {\mathbf{b}}_{j}
-{1 \over 2} \delta^{i}_{j} N_{b}
\en
and the conformal hamiltonian $E$ is simply
\eq
E = {1 \over 2} (N_a + N_b + 2P),
\en
where $N_{a} \equiv {\mathbf{a}}^{i} \cdot {\mathbf{a}}_{i}$,
$N_{b} \equiv {\mathbf{b}}^{j} \cdot {\mathbf{b}}_{j}$ are the
bosonic number operators.

First step in constructing a  positive energy UIR is
to identify a set of states in the Fock space transforming in an irreducible representation $|\Omega (j_L,j_R,E )\rangle$ of
$SU(2)_{L}\times SU(2)_{R}\times U(1)_{E}$ where
 $(j_{L},j_{R})$ are $SU(2)_L \times SU(2)_R$ spins that are annihilated by all the operators
${\mathbf{a}}_{i}\cdot {\mathbf{b}}_{j}$ of  subspace $\mathcal{L}^{-}$:
\eq
\mathbf{a}_{i}\cdot\mathbf{b}_{j} |\Omega (j_L,j_R,E )\rangle =0 \,\,\,\, \forall \,\,  \mathbf{a}_{i}\cdot\mathbf{b}_{j} \in \mathcal{L}^-
\en
Acting  repeatedly with the di-creation operators
$\mathbf{a}^{i}\cdot\mathbf{b}^{j}$
of $\mathcal{L}^{+}$
on $|\Omega\rangle$, one generates the basis of a positive energy UIR of the
full  group $SU(2,2)$:
\eq
|\Omega (j_L,j_R,E )\rangle \,\, , \, \mathcal{L}^+ |\Omega (j_L,j_R,E )\rangle \,\,, \, \mathcal{L}^+ \mathcal{L}^+ |\Omega (j_L,j_R,E )\rangle \,\, ,\cdots
\en
We shall refer to $|\Omega (j_L,j_R,E )\rangle $ as the lowest energy irreducible representation of $ SU(2)\times SU(2)\times U(1)$.

\section{Six-dimensional $\Gamma$ matrices}
\def\theequation{B.\arabic{equation}}
\setcounter{equation}0

Our choice of $\Gamma$-matrices is given below:
\eqn
\Gamma_{0} & = & \sigma_{3} \otimes I_{2} \otimes I_{2}, \nn \\
\Gamma_{1} & = & i \sigma_{1} \otimes \sigma_{2} \otimes I_{2}, \nn \\
\Gamma_{2} & = & i \sigma_{1} \otimes \sigma_{1} \otimes \sigma_{2}, \nn \\
\Gamma_{3} & = & i \sigma_{1} \otimes \sigma_{3} \otimes \sigma_{2}, \nn \\
\Gamma_{4} & = & i \sigma_{2} \otimes I_{2} \otimes \sigma_{2} \nn, \\
\Gamma_{5} & = & i \sigma_{2} \otimes \sigma_{2} \otimes \sigma_{1}, \enn where
$\sigma_{1}$, $\sigma_{2}$ and $\sigma_{3}$ are the Pauli matrices and $I_{2}$ is  the $2
\times 2$ identity matrix.

Therefore, \eq \Gamma_{7} = - \Gamma_{0} \Gamma_{1} \Gamma_{2}
\Gamma_{3} \Gamma_{4} \Gamma_{5} = - \sigma_{2} \otimes \sigma_{2} \otimes \sigma_{3}.
\en

The $\Sigma$ matrices in $d = 6$, the analogs of Pauli matrices
$\sigma_{\mu}$ in $d = 4$,  are given below.
\eqn
\left( \Sigma_{0} \right) & = & - i \sigma_{2} \otimes \sigma_{3},
  \\
\left( \Sigma_{1} \right) & = & i \sigma_{2} \otimes I_{2}, \\
\left( \Sigma_{2} \right) & = & i \sigma_{1} \otimes \sigma_{2},
\\
\left( \Sigma_{3} \right) & = & i \sigma_{3} \otimes \sigma_{2},
\\
\left( \Sigma_{4} \right) & = & I_{2} \otimes \sigma_{2},\\
\left( \Sigma_{5} \right) & = & \sigma_{2} \otimes \sigma_{1}
\enn


\end{document}